\documentclass[reprint,sort&compress, nobibnotes, floatfix, aps,  prl]{revtex4-2}
\usepackage{hyperref}
\usepackage[vietnamese, english]{babel}
\usepackage{amssymb}
\usepackage{amsmath}
\usepackage{graphicx}
\usepackage{color}
\hypersetup{
     colorlinks   = true,
     citecolor    = blue
}
\usepackage{physics}
\usepackage{float}
\usepackage{bbold}
\usepackage[normalem]{ulem}
\renewcommand{\vec}[1]{\boldsymbol{#1}}
\graphicspath{{Figs/}}
\newcommand{\be}{\begin{equation}}
\newcommand{\ee}{\end{equation}}
\newcommand{\bea}{\begin{eqnarray}}
\newcommand{\eea}{\end{eqnarray}}

\def\nn{\nonumber}

\def\pref{\eqref}

\begin{document}

\title{Superconductivity from Repulsive Interactions in Rhombohedral Trilayer Graphene:
a Kohn-Luttinger-Like Mechanism}

\author{Tommaso Cea}
\affiliation{Imdea Nanoscience, Faraday 9, 28015 Madrid, Spain}
\author{Pierre A. Pantale\'on}
\affiliation{Imdea Nanoscience, Faraday 9, 28015 Madrid, Spain}
\author{\foreignlanguage{vietnamese}{Võ Tiến Phong}}
\affiliation{Department of Physics and Astronomy, University of Pennsylvania, Philadelphia PA 19104}
\author{Francisco Guinea}
\affiliation{Imdea Nanoscience, Faraday 9, 28015 Madrid, Spain}
\affiliation{Donostia International Physics Center, Paseo Manuel de Lardiz\'abal 4, 20018 San Sebasti\'an, Spain}
\affiliation{ Ikerbasque. Basque Foundation for Science. 48009 Bilbao. Spain. }

\date{\today}

\begin{abstract}
We study the emergence of superconductivity in rhombohedral trilayer graphene due purely to the long-range Coulomb repulsion. This repulsive-interaction-driven phase in rhombohedral trilayer graphene is significantly different from those found in twisted bilayer and trilayer graphenes. In the latter case, the nontrivial momentum-space geometry of the Bloch wavefunctions leads to an effective attractive electron-electron interaction; this allows for less modulated order parameters and for spin-singlet pairing. In rhombohedral trilayer graphene, we instead find spin-triplet superconductivity with critical temperatures up to $0.15$ K. The critical temperatures strongly depend on electron filling and peak where the density of states diverge. The order parameter shows a significant modulation within each valley pocket of the Fermi surface.
\end{abstract}

\maketitle

{\it Introduction $-$ } Recently, superconductivity was experimentally observed in a three-layer graphene stack with rhombohedral (ABC) arrangement that is tunable by an applied interlayer bias~\cite{ZXTWY21}. Following this, several theories have been proposed to account for the onset of effective attractive interaction between electrons mediated by different pairing mechanisms:  electron-phonon coupling~\cite{chou_dassarma_cm21}, spin fluctuations near an antiferromagnetic phase~\cite{dai_cm21,dong_cm21}, direct coupling by the screened Coulomb interaction~\cite{GHMB21}, or pairing mediated by the proximity to a correlated insulator~\cite{CWBZ21,dong_cm21}. Although different in details, all of these proposals made use of the fact that the density of states (DOS) of ABC trilayer graphene near charge neutrality can be greatly enhanced by applying a gate voltage across the three layers. Ignoring possible weak spin-orbit couplings, intrinsic ABC trilayer graphene is a semimetal with an approximate cubic band degeneracy at the zone corners \cite{M69,DD02,AG08,ZSMM10,koshino_prb10,Betal11,KHV11,KHH13,Letal14,PBM17,Cetal19,Letal19,yin_prl19,chittari_prl19,Chen2019,Chen2020,Setal20,zhou_cm21_bis}.  When resolved close to these points, the cubic degeneracy actually splits into three Dirac cones, creating a trigonally-warped Fermi surface. As a perpendicular electric field is applied, inversion symmetry is broken, and these Dirac points acquire a finite mass. As a result, the local band dispersion can be nearly quenched, generating a van Hove singularity that favors the emergence of correlated electronic phases.

Similar physics can also be found in three-dimensional (3D) rhombohedral graphite, which is a nodal line semimetal that has a flat electronic band at the top and bottom surfaces of a sufficiently wide stack~\cite{AMV18}. The associated divergent DOS is expected to enhance electron-electron interactions and  lead to broken-symmetry phases, including superconductivity~\cite{KHV11,KHH13} and magnetism~\cite{PBM17}. Experimentally, gaps and broken-symmetry phases in finite rhombohedral stacks have been reported~\cite{Letal14,Cetal19,Letal19,yin_prl19,Chen2019,Chen2020,Setal20,zhou_cm21_bis}. The partially flat bands in rhombohedral stacks make these systems spectrally similar to magic-angle twisted bilayer graphene~\cite{Cao2018,Cao2018_bis}, importantly without the need for a superlattice structure.

\begin{figure}
    \centering
    \includegraphics[scale=0.31]{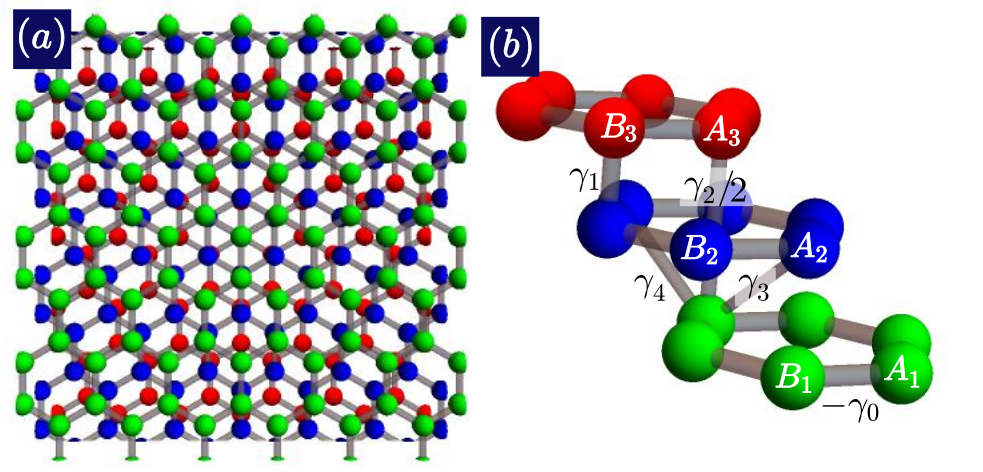}
    \caption{\textbf{Lattice structure of ABC RTG.} (a) The 3D crystal structure viewed from the top. (b) Some representative hopping parameters between carbon atoms.}
    \label{fig:crystalstructur}
\end{figure}

Inspired by these observations, we analyze here the appearance of superconductivity in rhombohedral trilayer graphene (RTG). We assume that the only electron-electron coupling is via the long-range Coulomb interaction. We analyze the possibility of pairing using a diagrammatic technique, similar to the Kohn-Luttinger approach~\cite{KL_prl65} to superconductivity due to repulsive interactions (see also~\cite{GHMB21}). The same scheme has been already applied to twisted bilayer graphene~\cite{CG21}, and to twisted trilayer graphene~\cite{phong_cm21}. The use of the same technique allows us to compare the emergence of superconductivity in twisted and rhombohedral stacks.  As discussed below, the calculation leads to superconducting (SC) phases in both types of materials, although the physical origin of superconductivity and the superconducting order parameter (OP) are significantly different in the two cases.

{\it Tight-binding Hamiltonian $-$} The 3D crystal structure of  RTG is shown in Fig. \ref{fig:crystalstructur}(a). Each unit cell consists of six carbon atoms, two per layer, connected to each other via hopping amplitudes as shown in Fig. \ref{fig:crystalstructur}(b). The minimal tight-binding Hamiltonian is given by \cite{zhang_prb10}
\begin{widetext}
\bea\label{H0_mat}
H_{tb}(\vec{k})=
\begin{pmatrix}
\Delta_1+\Delta_2&-\gamma_0u(\vec{k})&\gamma_4u^*(\vec{k})&\gamma_1&0&0\\
-\gamma_0u^*(\vec{k})&\Delta_1+\Delta_2+\delta&\gamma_3v(\vec{k})&\gamma_4u^*(\vec{k})&
\gamma_2/2&0\\
\gamma_4u(\vec{k})&\gamma_3v^*(\vec{k})&-2\Delta_2&-\gamma_0u(\vec{k})&
\gamma_4v^*(\vec{k})&\gamma_1\\
\gamma_1&\gamma_4u(\vec{k})&-\gamma_0u^*(\vec{k})&-2\Delta_2&\gamma_3u(\vec{k})&
\gamma_4u^*(\vec{k})\\
0&\gamma_2/2&\gamma_4v(\vec{k})&\gamma_3u^*(\vec{k})&\Delta_2-\Delta_1+\delta&
-\gamma_0v(\vec{k})\\
0&0&\gamma_1&\gamma_4u(\vec{k})&-\gamma_0v^*(\vec{k})&\Delta_2-\Delta_1
\end{pmatrix},
\eea
\end{widetext}
where $\gamma_i$ are the hopping amplitudes, $\Delta_1$ is a potential difference between nearest neighbor layers which takes into account an external displacement field, $\Delta_2$ is the potential difference between the middle layer compared to mean potential of the outer layers, $\delta$ encodes an on-site potential which is only present at sites $B_1$ and $A_3$ since these two atoms do not have a neighbor on the middle layer, and $u(\vec{k})=1+2\cos\left(k_xa/2\right)e^{-ik_ya\sqrt{3}/2}$, $v(\vec{k})=e^{ik_ya\sqrt{3}}u(\vec{k})$, with $a=2.46$ {\AA} is the lattice constant of graphene. The optimal values of the minimal tight-binding parameters, $\gamma_i$ and $\delta$, have been reported in the  literature~\cite{koshino_prb10,zhang_prb10,zibrov_prl18,yin_prl19,chittari_prl19,shi_nat20}. Here we use the ones calculated in Refs. \cite{zibrov_prl18, zhou_cm21_bis}. These parameters are tabulated in Table \ref{table:ABC_parameters}.
\begin{table}
\begin{center}
\begin{tabular}{| c c c c c c c|}
\hline
$\gamma_0$&$\gamma_1$&$\gamma_2$&$\gamma_3$&$\gamma_4$&$\delta$&$\Delta_2$\\
\hline
3.1		    &0.38              &-0.015           &0.29              &0.141            &-0.0105 &-0.0023\\
 \hline
\end{tabular}
\end{center}
\caption{Minimal tight-binding parameters of RTG, expressed in eV (see also~\cite{zibrov_prl18,zhou_cm21_bis}).} 
\label{table:ABC_parameters}
\end{table}

\begin{figure}
\includegraphics[width=2.5in]{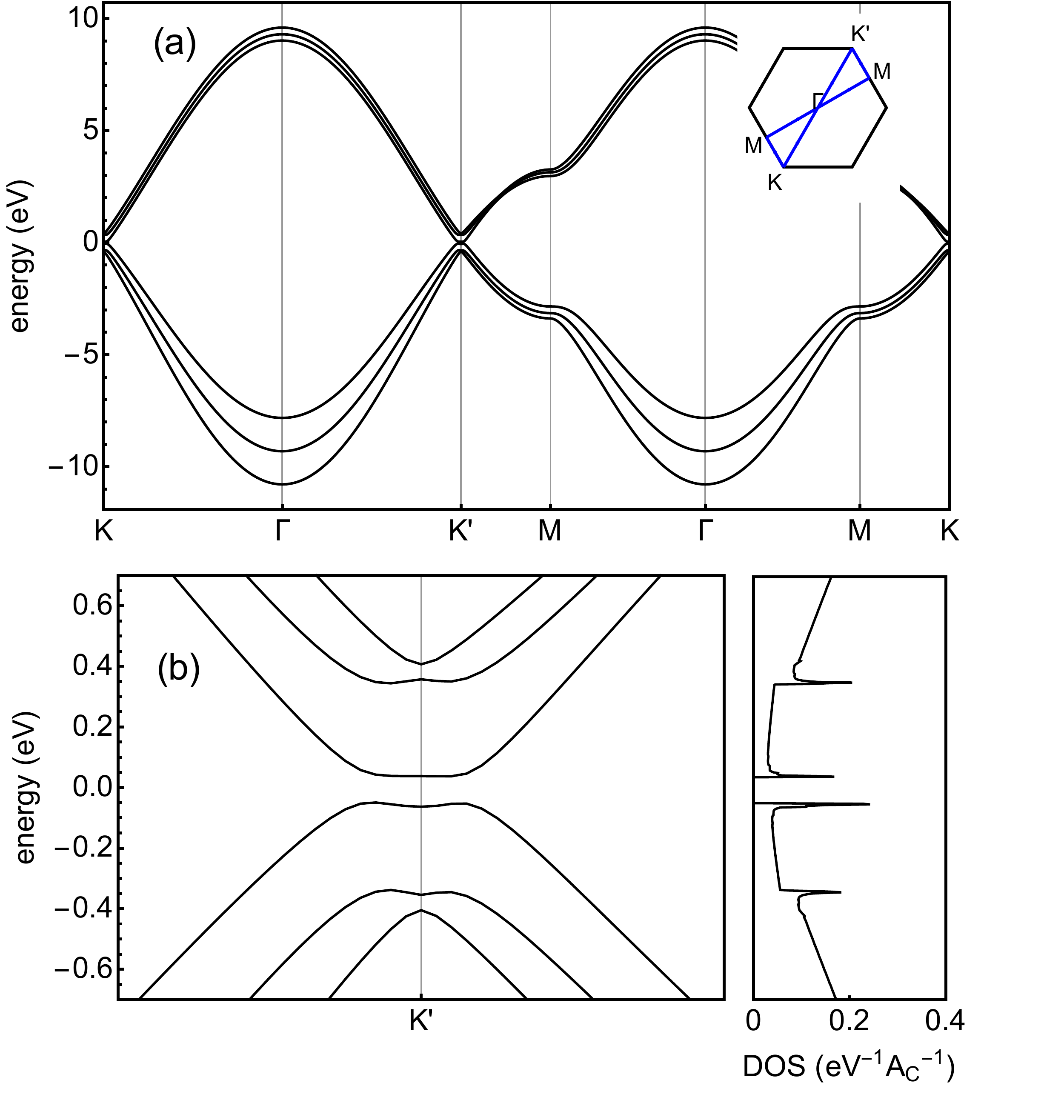}
\caption{ \textbf{Band structure of RTG.} (a) Full band structure calculated for $\Delta_1=50$ meV. The inset shows the BZ and the high-symmetry path used for calculation.
(b) Left: detail of the band structure close to  $K'$. 
(b) Right: DOS.
}
\label{fig:band_structure}
\end{figure}

The single-particle band structure corresponding to  model \pref{H0_mat} for $\Delta_1=50$ meV along a high-symmetry path of the Brillouin zone (BZ) is shown in Fig. \ref{fig:band_structure}(a). Since we are only interested in the lightly-doped regime, only the local band structure around $E=0$ is of relevance to us. This region in momentum space is shown in Fig. \ref{fig:band_structure}(b) along with the corresponding DOS expressed in units of eV$^{-1}A_c^{-1}$, where $A_c=\sqrt{3}a^2/2$ is the area of one unit cell. We draw attention to the sizeable gap at the charge neutrality point (CNP) generated by the external electric field, and to the van Hove singularities at the gap's edges due to band flattening at $K$ and $K'.$ See the Supplementary Information for more details on the non interacting band structure.

{\it Long-range Coulomb interaction and internal screening $-$} To account for electron-electron interactions, we assume that two electrons separated by a distance $r$ experience a $r^{-2}$ long-range Coulomb repulsion
\bea\label{V0_r}
\begin{matrix}
V_C(r)=&\frac{e^2}{4\pi\epsilon_0\epsilon r}&\text{ , for }r\ne0,\\
V_C(0)=&\frac{w_0}{\epsilon},&
\end{matrix}
\eea
where $e$ is the electron charge, $\epsilon_0$ is the dielectric constant of  vacuum, and $\epsilon$ is the relative dielectric constant of the environment. In this work, we set $\epsilon=4$, which reproduces accurately the screening by a substrate of hexagonal boron nitride (hBN). The parameter $w_0$ accounts for the local repulsion, which we set to $w_0=17$ eV following Ref.~\cite{wehling_prl11}. As the potential $V_C$ varies slowly on the atomic scale, we approximate the interaction between two electrons as only depending on the distance between the centers of the two unit cells in which the electrons reside. In  reciprocal space,  $V_C$ is given by:
\bea
V_C(\vec{q})=\sum_{\vec{R}}V_C\left(|\vec{R}|\right)e^{-i\vec{q}\cdot\vec{R}},
\eea
where $\vec{q}\in$ BZ,  the sum runs over all  positions $\vec{R}$ of the lattice, with  periodic boundary condition imposed by the finite grid used to sample the BZ.

\begin{figure}
\includegraphics[width=2.8in]{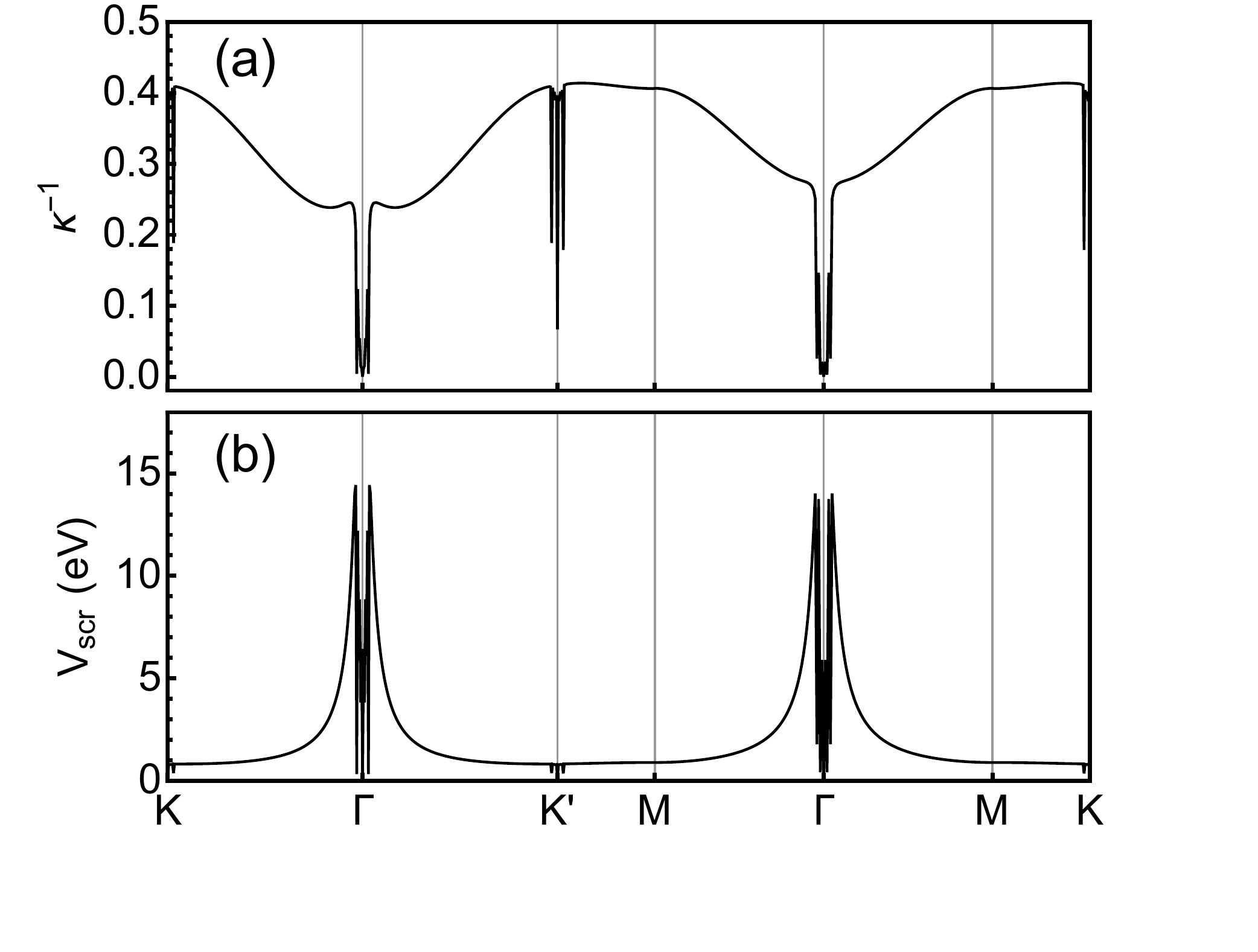}
\caption{ \textbf{Screened Coulomb potential.} (a) Inverse of the dielectric function, $\kappa^{-1}(\vec{q})=\left[1-V_C(\vec{q})\Pi(\vec{q})\right]^{-1}$. (b) Screened potential, $V_{scr}(\vec{q})$.  The profiles are computed along the high-symmetry path shown in  Fig. \ref{fig:band_structure}(a), and obtained for $\Delta_1=50$ meV and electronic density $n_e=-1.91\times10^{12}$ cm$^{-2}$. }
\label{fig:V_RPA}
\end{figure}

In order to describe  internal screening due to  particle-hole excitations, we use the static random phase approximation (RPA), leading to the usual renormalization of $V_C$
\bea\label{V_RPA}
V_{scr}(\vec{q})=\frac{V_C(\vec{q})}{1-V_C(\vec{q})\Pi(\vec{q})},
\eea
where $\Pi(\vec{q})$ is the zero-frequency limit of the charge susceptivity,
as given by
\bea\label{PI_q}
\Pi(\vec{q})&=&\frac{2}{N_c}\sum_{\vec{k}nm}
\frac{f(\xi_{n,\vec{k}})-f(\xi_{m,\vec{k}+\vec{q}})}
{\epsilon_{n,\vec{k}}-\epsilon_{m,\vec{k}+\vec{q}}}\times\\
&\times&
\left|\left\langle
\vec{\psi}_{m,\vec{k}+\vec{q}}|\vec{\psi}_{n,\vec{k}}
\right\rangle\right|^2,\nn
\eea
where $N_c$ is the number of unit cells,  $\epsilon_{n,\vec{k}}$ is the $n$-th band energy at wavevector $\vec{k}$, $\vec{\psi}_{n,\vec{k}}$ is the corresponding six-component eigenvector, $f(\xi)=\left[1+e^{\xi/(K_BT)}\right]^{-1}$ is the Fermi-Dirac distribution at the temperature $T$, $\xi_{n,\vec{k}}=\epsilon_{n,\vec{k}}-\mu,$ and $\mu$ is the chemical potential. The factor of two in front of  Eq. \pref{PI_q} accounts for  spin degeneracy. As an example, Fig.~\ref{fig:V_RPA} shows (a) the profile of the inverse of the dielectric function, $\kappa^{-1}(\vec{q})=\left[1-V_C(\vec{q})\Pi(\vec{q})\right]^{-1}$, and (b) of the screened potential, $V_{scr}(\vec{q})$, computed along the high-symmetry path of the BZ shown in  Fig. \ref{fig:band_structure}(a), and obtained for $\Delta_1=50$ meV and electronic density $n_e=-1.91\times10^{12}$ cm$^{-2}$. To perform the calculation, we used $N_c=12\times10^4$, which is enough to finely resolve the band structure close to the Fermi surface (FS). The results display an overall strong screening. Remarkably, $V_{scr}(\vec{q})$ vanishes at the centre of the BZ, the point $\Gamma$, which means that $\Pi(\vec{q})$ diverges as $\vec{q}\to0$ and implies that $V_{scr}$ is locally attractive in real space. 

{\it  Superconductivity $-$} Next, we assume that the interaction which leads to pairing in RTG is the long-range Coulomb interaction (Ref.~\cite{GHMB21} makes the same assumption). 
The calculations carried out in Ref.~\cite{CG21,phong_cm21} include, for completeness, the coupling of electronic charge oscillations to longitudinal phonons, as these phonons modify the screening of the Coulomb interaction. It is interesting to note that the inclusion of longitudinal phonons does not change significantly the results reported here.

The critical temperature for the onset of superconductivity in RTG can be obtained from the linearized gap equation
\bea
\label{lin_gap_eq}
\Delta_{ij}(\vec{k})&=-\frac{K_BT}{N_c}\sum_{\vec{k}'\omega}\sum_{i'j'}
V_{scr}(\vec{k}-\vec{k}') \times \nonumber \\ &\times G_{ii'}(\vec{k}',i\hbar\omega)G_{jj'}(-\vec{k}',-i\hbar\omega)\Delta_{i'j'}(\vec{k}'),
\eea
where $\omega$ are fermionic Matsubara frequencies, $i,i',j,j'$ label the sub-lattice/layer
degree of freedom, and $G_{ii'}(\vec{k},i\hbar\omega)$ is the normal-state single-particle Green's function
\bea
G_{ii'}(\vec{k},i\hbar\omega)=
\sum_n\frac{\psi^i_{n,\vec{k}}\psi^{i',*}_{n,\vec{k}}}{i\hbar\omega-\xi_{n,\vec{k}}}.
\eea
Our framework is similar to the Kohn-Luttinger scheme~\cite{KL_prl65}. The approach in~\cite{KL_prl65} includes all processes up to second order in perturbation theory. Our approach neglects exchange-like diagrams, but, on the other hand, includes all bubble diagrams to infinite orders. The multiplicity of these diagrams is equal to the number of electron flavors, in the present case ${\cal N}_f = 2$. Hence, it can be considered an expansion in powers of ${\cal N}_f^{-1}$.

Upon projecting  Eq. \pref{lin_gap_eq} onto the band basis and performing the Matsubara sum, we rewrite it as
\bea\label{lin_gap_eq2}
\Delta_{m_1m_2}(\vec{k})=\sum_{\vec{k}'n_1n_2}\Gamma_{m_1m_2,n_1n_2}(\vec{k},\vec{k}')
\Delta_{n_1n_2}(\vec{k}'),
\eea
where
\bea
\Delta_{m_1m_2}(\vec{k})&=&
\sum_{ij}\psi^{i,*}_{m_1,\vec{k}}\psi^i_{m_2,\vec{k}}\Delta_{ij}(\vec{k})\times\\
&\times&
\sqrt{\frac{f(-\xi_{m_2,\vec{k}})-f(\xi_{m_1,\vec{k}})}{\xi_{m_2,\vec{k}}+\xi_{m_1,\vec{k}}}},
\nn
\eea
and $\Gamma_{m_1m_2,n_1n_2}(\vec{k},\vec{k}')$ is the Hermitian kernel
\bea\label{Gamma_kernel}
\begin{split}
& \Gamma_{m_1m_2,n_1n_2}(\vec{k},\vec{k}')\\
&= -\frac{1}{N_c}V_{scr}(\vec{k}-\vec{k}')
\left\langle\vec{\psi}_{m_1,\vec{k}}|\vec{\psi}_{n_1,\vec{k}'} \right\rangle
\left\langle\vec{\psi}_{n_2,\vec{k}'}|\vec{\psi}_{m_2,\vec{k}} \right\rangle\times\\
&\times
\sqrt{\frac{f(-\xi_{m_2,\vec{k}})-f(\xi_{m_1,\vec{k}})}{\xi_{m_2,\vec{k}}+\xi_{m_1,\vec{k}}}}
\sqrt{\frac{f(-\xi_{n_2,\vec{k}'})-f(\xi_{n_1,\vec{k}'})}{\xi_{n_2,\vec{k}'}+\xi_{n_1,\vec{k}'}}}.
\end{split}
\eea
We make use of the  time-reversal symmetry of Hamiltonian \pref{H0_mat}, which implies  $\xi_{n,-\vec{k}}=\xi_{n,\vec{k}}$ and $\vec{\psi}_{m,-\vec{k}}=\vec{\psi}^*_{m,\vec{k}}$. At a critical temperature, $T_c$, the largest eigenvalue of the kernel $\Gamma_{m_1m_2,n_1n_2}(\vec{k},\vec{k}')$ is 1. The corresponding eigenvector provides the symmetry of the OP.

We diagonalize numerically the Kernel of  Eq. \pref{Gamma_kernel}. As the leading contribution to Eq.~\pref{lin_gap_eq2} comes from the states closest to the FS, we cut off  phase space by considering only the states satisfying $|\xi_{n,\vec{k}}|\le w$, with $w=30$ meV. In order to rule out finite-size effects and to finely sample the FS at densities on the order of $\sim10^{12}$ cm$^{-2}$, we implement a length renormalization: $a\rightarrow a_s=s\times a$, where $s>1$ is the scale factor and $a_s$ is the effective lattice spacing. This procedure defines an effective tight-binding model where the hopping amplitudes $\gamma_0,$ $\gamma_3,$ and $\gamma_4$ are rescaled according to $\gamma_i\rightarrow\gamma_{i,s}=\gamma_i/s$, $i=0,3,4$. This procedure reduces the size of the BZ by a factor of $s^2,$ allowing us to study considerably larger meshes that would otherwise be numerically prohibitive. In doing so, we are able to obtain a finer momemtum resolution close to the CNP to improve accuracy.

\begin{figure}
\includegraphics[width=2.5in]{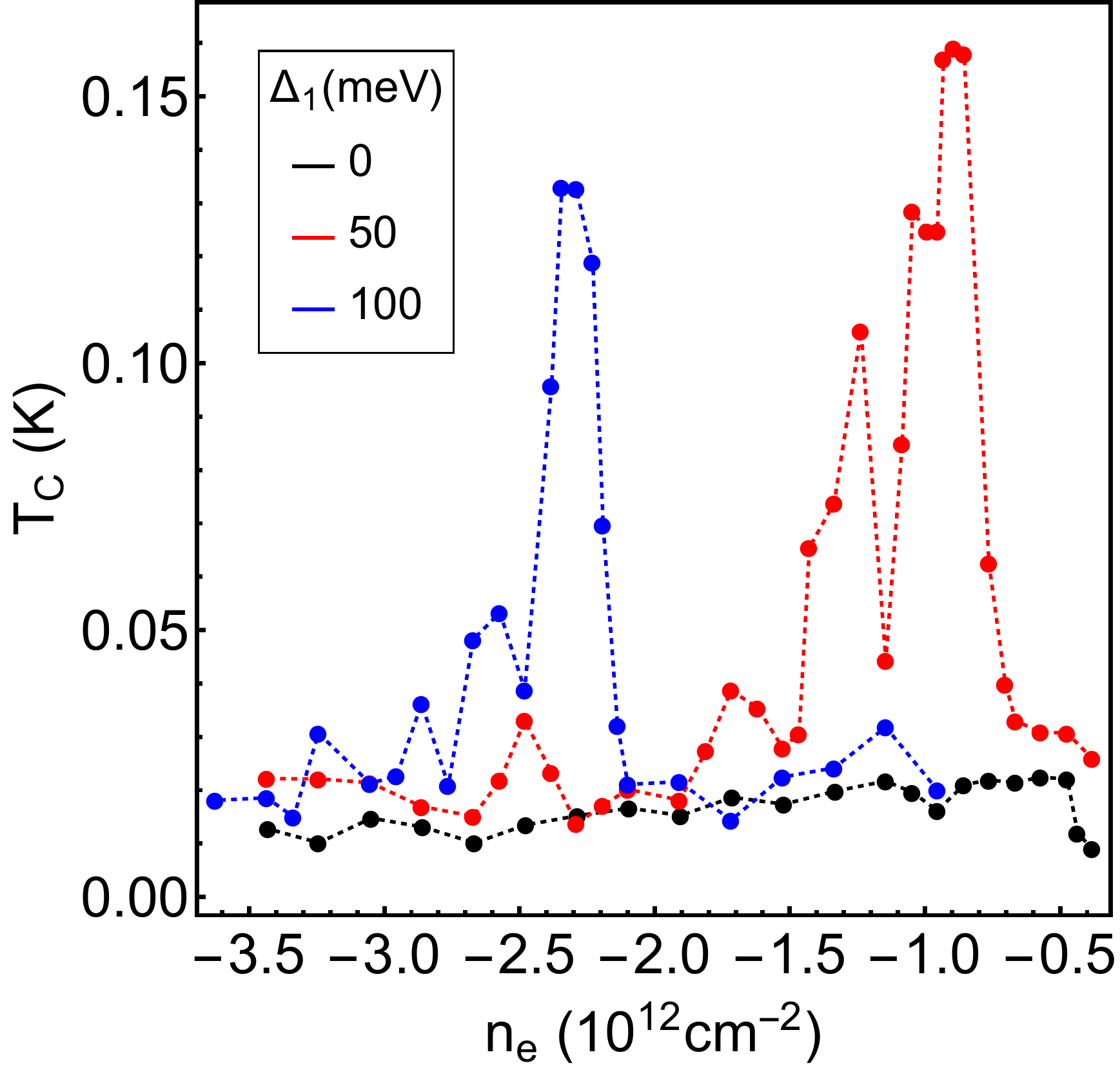}
\caption{
\textbf{Critical temperature as a function of filling for various values of the interlayer bias.}
}
\label{fig:Tc}
\end{figure}

\begin{figure}
\includegraphics[width=3.in]{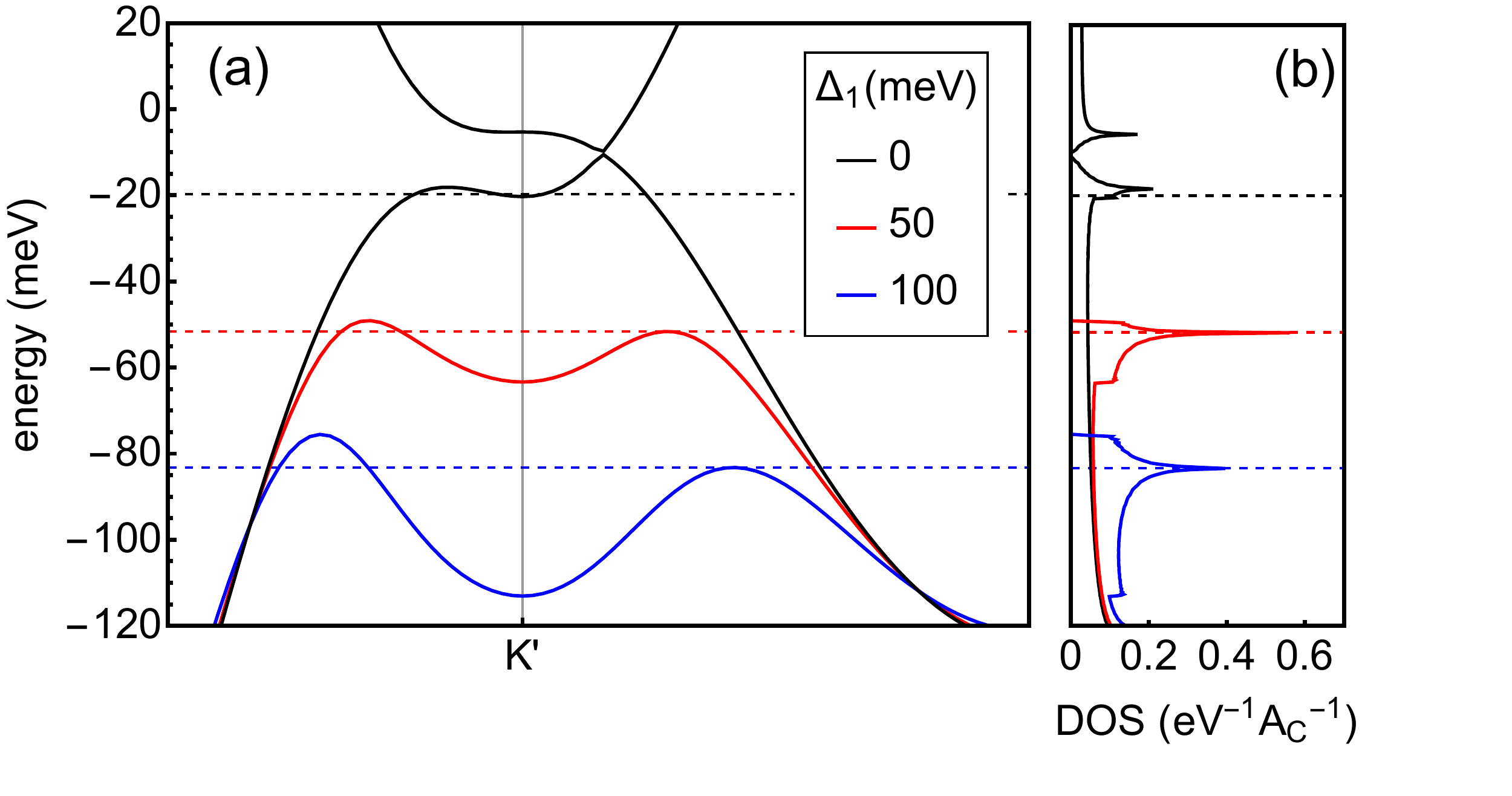}
\caption{ \textbf{Bias-induced van Hove singularities.}
Electronic bands (a) and DOS (b) close to the CNP, obtained for the values of $\Delta_1$ considered in  Fig. \ref{fig:Tc}. The horizontal dashed lines identify the Fermi levels corresponding to the values of $n_e$ which maximize $T_c$ in Fig. \ref{fig:Tc}.
}
\label{fig:bands_and_DOSs_Delta1}
\end{figure}

{\it Critical superconducting temperature $-$} Fig.~\ref{fig:Tc} shows the value of the critical temperature as a function of  electronic density, $n_e$, for $\Delta_1=0,$ $50,$ $100$ meV. The results are  obtained with a grid of $3\times10^4$ points in the BZ, upon rescaling with $s=10$, meaning that we are considering $3\times10^6$ unit cells of the atomic RTG. The critical temperatures feature pronounced maxima on the order of $0.1-0.2$ K for finite values of $\Delta_1$. In contrast, we do not observe any appreciable enhancement of $T_c$ without a bias.  To gain insight into this behavior, we show in  Fig.~\ref{fig:bands_and_DOSs_Delta1} the bands (a) and DOS (b) close to the CNP, obtained for the values of $\Delta_1$ considered in Fig. \ref{fig:Tc}. We observe that a finite bias significantly enhances the van Hove singularities near the band edge, a feature that is absent in the zero-bias limit. In Fig. \ref{fig:bands_and_DOSs_Delta1}, the horizontal dashed lines identify the Fermi levels corresponding to the values of $n_e$ which maximize $T_c$ in Fig. \ref{fig:Tc}. These Fermi energies match the position of the Van Hove singularities with great accuracy, showing that superconductivity is strongly enhanced when the Fermi level is close to a peak in the DOS. In addition, given a finite bias, a sizeable $T_c$ survives only in a narrow region of $n_e$ around an optimal value, thus providing a tool to trigger  superconductivity by tuning $n_e$ and/or $\Delta_1$.  
It is worth noting that the results reported in Fig. \ref{fig:Tc} are in reasonable agreement with the experimental data of  Ref.~\cite{ZXTWY21}, in terms of both the magnitude of the critical temperatures and the range of densities reported.

\begin{figure}
\includegraphics[width=1.6in]{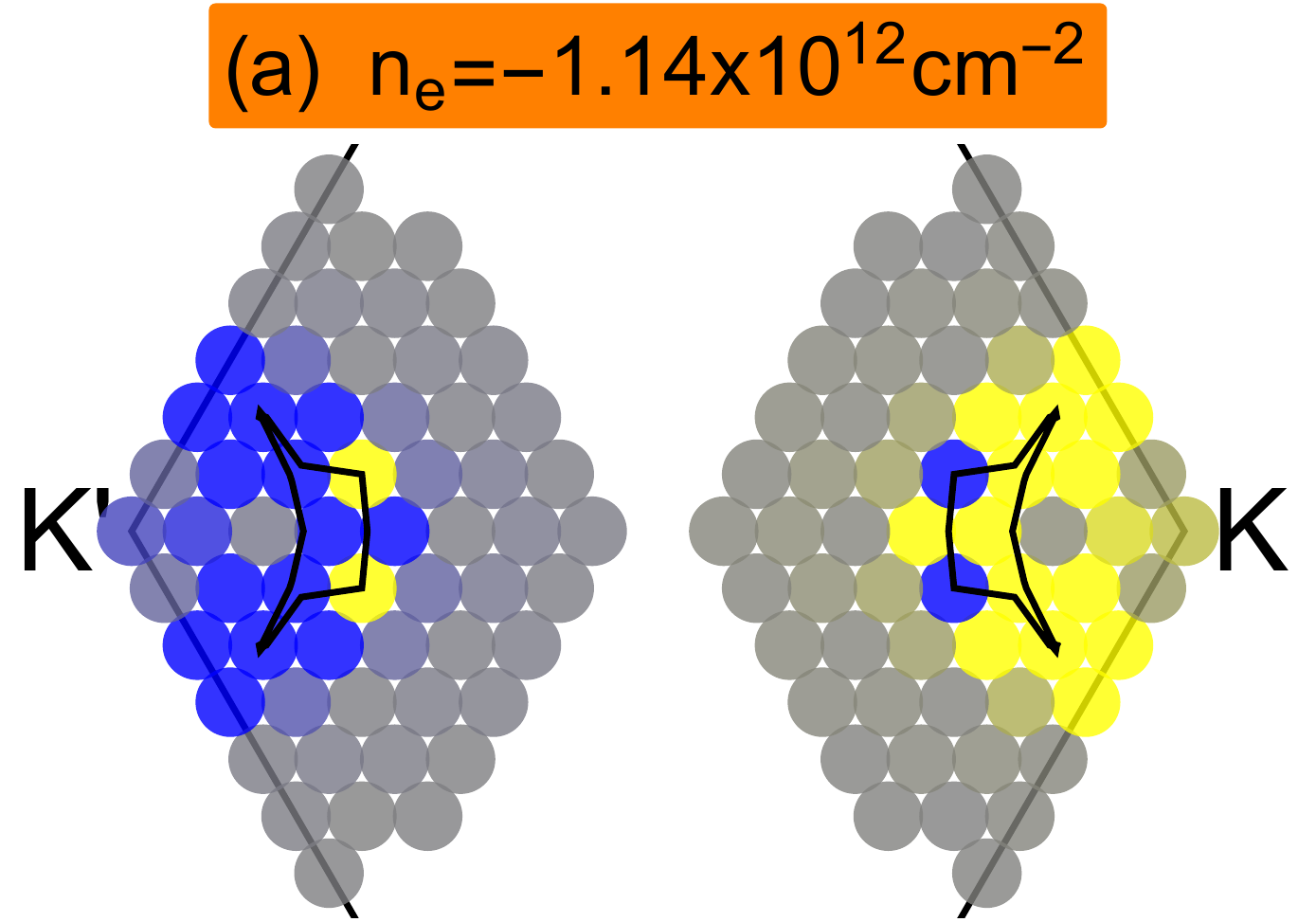}
\includegraphics[width=1.6in]{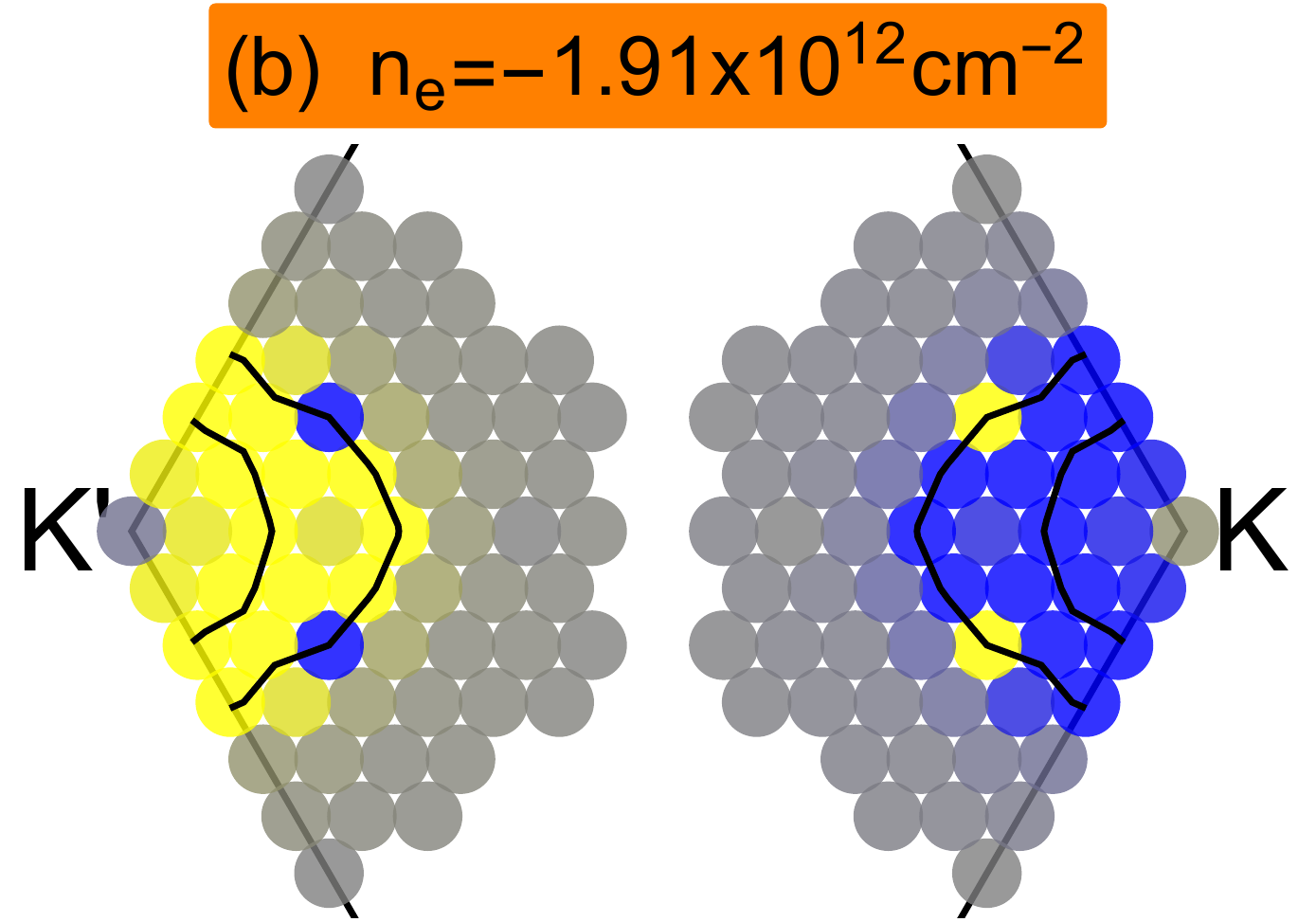}\\
\includegraphics[width=1.7in]{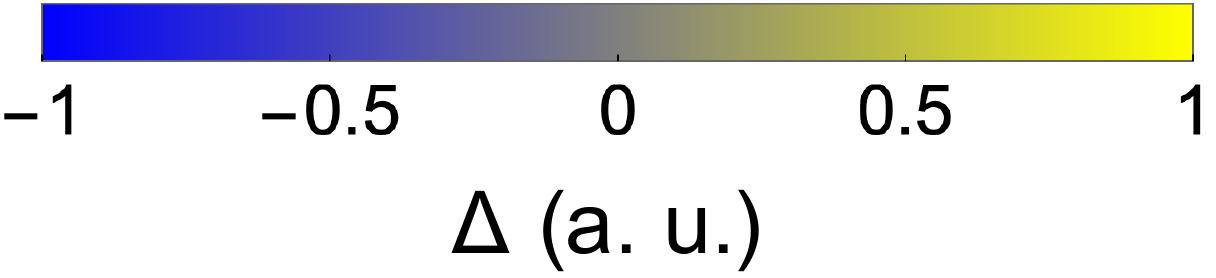}
\caption{
\textbf{Symmetry of the superconducting order parameter in the vicinity of  $K$ and $K'$.} These are calculated for $\Delta_1=50$ meV and $n_e=-1.14\times10^{12}$ cm$^{-2}$ (a) and $-1.91\times10^{12}$ cm$^{-2}$  (b). The black lines identify the FS. 
}
\label{fig:OP_symm}
\end{figure}

{\it Symmetry of the superconducting order parameter $-$} Finally, we study the superconducting order parameter. Fig. \ref{fig:OP_symm} shows the symmetry of the SC OP in the vicinity of $K$ and $K'$, obtained for $\Delta_1=50$ meV and $n_e=-1.14,-1.91\times10^{12}$ cm$^{-2}$, which are representative of most of the cases we have studied. The black lines identify the FS. These results have been obtained without scaling, $s=1$, by using $N_c=12\times10^4$ unit cells of the atomic RTG. As expected, the OP is nonzero only in a narrow region of the BZ around the FS, implying that only the electrons close to the Fermi level participate in  Cooper pairing. In addition, the OP clearly displays  $A_2$ symmetry, meaning that it is antisymmetric upon exchanging $\vec{k}\rightarrow-\vec{k}$. Because  Hamiltonian \pref{H0_mat} is spin degenerate and the interaction, Eq. \pref{V0_r}, does not couple different spin flavors, the gap equations \pref{lin_gap_eq} and \pref{lin_gap_eq2} do not contain explicitly the spin indices, implying that they cannot distinguish between spin-triplet and spin-singlet superconductivity. However,  $A_2$ symmetry necessarily implies that the Cooper pairs must be symmetric (i. e. triplet) in spin space in order for the total wavefunction to be antisymmetric upon exchanging the two electrons\cite{hund}. Finally, it is worth noting that the OP changes sign within each valley pocket of the FS. The pairing potential represented by the kernel $\Gamma_{m_1m_2,n_1n_2}(\vec{k},\vec{k}')$, Eq. \pref{Gamma_kernel}, is repulsive in  reciprocal space,  and the eigenvector corresponding to the eigenvalue $1$ cannot have a constant sign. This is a general feature of  weak-coupling superconductivity induced by  electronic interactions, where the superconducting wave function displays a high angular momentum, as is the case for $p$- or $f$-wave superconductivity.

{\it Conclusions $-$} We have analyzed diagrammatically the existence of superconductivity in RTG. We assume that the leading electron-electron interaction is  Coulomb repulsion. Our results show that this interaction is enough to induce superconductivity in RTG, although it cannot be excluded that other excitations can contribute~\cite{chou_dassarma_cm21,dai_cm21,CWBZ21,dong_cm21}. The large DOS in RTG at low fillings leads to  a significant screening of the interaction. The screened Coulomb repulsion induces superconductivity with critical temperatures upward $0.15$ K that depend strongly on electron filling and are correlated with peaks in the DOS. The OP fluctuates in sign within each valley, in agreement with the existence of a repulsive interaction at small momenta. Overall, the OP is antisymmetric in the BZ, so that the pairs must be spin triplets. The method used here has also been applied to the study of superconductivity in twisted bilayer and trilayer graphene~\cite{CG21,phong_cm21}. In those cases, however, the combination of Umklapp processes and the complexity of the wavefunctions turns the interaction attractive at small momenta. As a result, the OP does not change within individual pockets of the FS. The superconductivity can be spin singlet/valley triplet or spin triplet/valley singlet. The small momentum modulation of the OP implies that long-range disorder is pair breaking in RTG, while that is not the case in twisted bilayer/trilayer graphene.

{\it Acknowledgements.}
This work was supported by funding from the European Commision, under the Graphene Flagship, Core 3, grant number 881603, and by the grants NMAT2D (Comunidad de Madrid, Spain),  SprQuMat and SEV-2016-0686, (Ministerio de Ciencia e Innovación, Spain). VTP acknowledges support from the NSF Graduate Research Fellowships Program and the P.D. Soros Fellowship for New Americans.


\bibliography{Literature}

\begin{thebibliography}{35}%
\makeatletter
\providecommand \@ifxundefined [1]{%
 \@ifx{#1\undefined}
}%
\providecommand \@ifnum [1]{%
 \ifnum #1\expandafter \@firstoftwo
 \else \expandafter \@secondoftwo
 \fi
}%
\providecommand \@ifx [1]{%
 \ifx #1\expandafter \@firstoftwo
 \else \expandafter \@secondoftwo
 \fi
}%
\providecommand \natexlab [1]{#1}%
\providecommand \enquote  [1]{``#1''}%
\providecommand \bibnamefont  [1]{#1}%
\providecommand \bibfnamefont [1]{#1}%
\providecommand \citenamefont [1]{#1}%
\providecommand \href@noop [0]{\@secondoftwo}%
\providecommand \href [0]{\begingroup \@sanitize@url \@href}%
\providecommand \@href[1]{\@@startlink{#1}\@@href}%
\providecommand \@@href[1]{\endgroup#1\@@endlink}%
\providecommand \@sanitize@url [0]{\catcode `\\12\catcode `\$12\catcode
  `\&12\catcode `\#12\catcode `\^12\catcode `\_12\catcode `\%12\relax}%
\providecommand \@@startlink[1]{}%
\providecommand \@@endlink[0]{}%
\providecommand \url  [0]{\begingroup\@sanitize@url \@url }%
\providecommand \@url [1]{\endgroup\@href {#1}{\urlprefix }}%
\providecommand \urlprefix  [0]{URL }%
\providecommand \Eprint [0]{\href }%
\providecommand \doibase [0]{https://doi.org/}%
\providecommand \selectlanguage [0]{\@gobble}%
\providecommand \bibinfo  [0]{\@secondoftwo}%
\providecommand \bibfield  [0]{\@secondoftwo}%
\providecommand \translation [1]{[#1]}%
\providecommand \BibitemOpen [0]{}%
\providecommand \bibitemStop [0]{}%
\providecommand \bibitemNoStop [0]{.\EOS\space}%
\providecommand \EOS [0]{\spacefactor3000\relax}%
\providecommand \BibitemShut  [1]{\csname bibitem#1\endcsname}%
\let\auto@bib@innerbib\@empty
\bibitem [{\citenamefont {{Zhou}}\ \emph
  {et~al.}(2021{\natexlab{a}})\citenamefont {{Zhou}}, \citenamefont {{Xie}},
  \citenamefont {{Taniguchi}}, \citenamefont {{Watanabe}},\ and\ \citenamefont
  {{Young}}}]{ZXTWY21}%
  \BibitemOpen
  \bibfield  {author} {\bibinfo {author} {\bibfnamefont {H.}~\bibnamefont
  {{Zhou}}}, \bibinfo {author} {\bibfnamefont {T.}~\bibnamefont {{Xie}}},
  \bibinfo {author} {\bibfnamefont {T.}~\bibnamefont {{Taniguchi}}}, \bibinfo
  {author} {\bibfnamefont {K.}~\bibnamefont {{Watanabe}}},\ and\ \bibinfo
  {author} {\bibfnamefont {A.~F.}\ \bibnamefont {{Young}}},\ }\bibfield
  {title} {\bibinfo {title} {Superconductivity in rhombohedral trilayer
  graphene},\ }\href
  {https://doi.org/https://doi.org/10.1038/s41586-021-03926-0} {\bibfield
  {journal} {\bibinfo  {journal} {Nature}\ } (\bibinfo {year}
  {2021}{\natexlab{a}})}\BibitemShut {NoStop}%
\bibitem [{\citenamefont {{Chou}}\ \emph {et~al.}(2021)\citenamefont {{Chou}},
  \citenamefont {{Wu}}, \citenamefont {{Sau}},\ and\ \citenamefont {{Das
  Sarma}}}]{chou_dassarma_cm21}%
  \BibitemOpen
  \bibfield  {author} {\bibinfo {author} {\bibfnamefont {Y.-Z.}\ \bibnamefont
  {{Chou}}}, \bibinfo {author} {\bibfnamefont {F.}~\bibnamefont {{Wu}}},
  \bibinfo {author} {\bibfnamefont {J.~D.}\ \bibnamefont {{Sau}}},\ and\
  \bibinfo {author} {\bibfnamefont {S.}~\bibnamefont {{Das Sarma}}},\
  }\bibfield  {title} {\bibinfo {title} {{Acoustic-phonon-mediated
  superconductivity in rhombohedral trilayer graphene}},\ }\href@noop {}
  {\bibfield  {journal} {\bibinfo  {journal} {arXiv e-prints}\ ,\ \bibinfo
  {eid} {arXiv:2106.13231}} (\bibinfo {year} {2021})},\ \Eprint
  {https://arxiv.org/abs/2106.13231} {arXiv:2106.13231 [cond-mat.supr-con]}
  \BibitemShut {NoStop}%
\bibitem [{\citenamefont {{Dai}}\ \emph {et~al.}(2021)\citenamefont {{Dai}},
  \citenamefont {{Hou}}, \citenamefont {{Zhang}}, \citenamefont {{Liang}},\
  and\ \citenamefont {{Ma}}}]{dai_cm21}%
  \BibitemOpen
  \bibfield  {author} {\bibinfo {author} {\bibfnamefont {H.}~\bibnamefont
  {{Dai}}}, \bibinfo {author} {\bibfnamefont {J.}~\bibnamefont {{Hou}}},
  \bibinfo {author} {\bibfnamefont {X.}~\bibnamefont {{Zhang}}}, \bibinfo
  {author} {\bibfnamefont {Y.}~\bibnamefont {{Liang}}},\ and\ \bibinfo {author}
  {\bibfnamefont {T.}~\bibnamefont {{Ma}}},\ }\bibfield  {title} {\bibinfo
  {title} {{Mott insulating state and d +i d superconductivity in an ABC
  graphene trilayer}},\ }\href {https://doi.org/10.1103/PhysRevB.104.035104}
  {\bibfield  {journal} {\bibinfo  {journal} {\prb}\ }\textbf {\bibinfo
  {volume} {104}},\ \bibinfo {eid} {035104} (\bibinfo {year} {2021})},\ \Eprint
  {https://arxiv.org/abs/2009.14647} {arXiv:2009.14647 [cond-mat.str-el]}
  \BibitemShut {NoStop}%
\bibitem [{\citenamefont {{Dong}}\ and\ \citenamefont
  {{Levitov}}(2021)}]{dong_cm21}%
  \BibitemOpen
  \bibfield  {author} {\bibinfo {author} {\bibfnamefont {Z.}~\bibnamefont
  {{Dong}}}\ and\ \bibinfo {author} {\bibfnamefont {L.}~\bibnamefont
  {{Levitov}}},\ }\bibfield  {title} {\bibinfo {title} {{Superconductivity in
  the vicinity of an isospin-polarized state in a cubic Dirac band}},\
  }\href@noop {} {\bibfield  {journal} {\bibinfo  {journal} {arXiv e-prints}\
  ,\ \bibinfo {eid} {arXiv:2109.01133}} (\bibinfo {year} {2021})},\ \Eprint
  {https://arxiv.org/abs/2109.01133} {arXiv:2109.01133 [cond-mat.supr-con]}
  \BibitemShut {NoStop}%
\bibitem [{\citenamefont {Ghazaryan}\ \emph {et~al.}(2021)\citenamefont
  {Ghazaryan}, \citenamefont {Holder}, \citenamefont {Maksym},\ and\
  \citenamefont {Berg}}]{GHMB21}%
  \BibitemOpen
  \bibfield  {author} {\bibinfo {author} {\bibfnamefont {A.}~\bibnamefont
  {Ghazaryan}}, \bibinfo {author} {\bibfnamefont {T.}~\bibnamefont {Holder}},
  \bibinfo {author} {\bibfnamefont {S.}~\bibnamefont {Maksym}},\ and\ \bibinfo
  {author} {\bibfnamefont {E.}~\bibnamefont {Berg}},\ }\bibfield  {title}
  {\bibinfo {title} {Unconventional superconductivity in systems with annular
  fermi surfaces: Application to rhombohedral trilayer graphene},\ }\href@noop
  {} {\  (\bibinfo {year} {2021})}\BibitemShut {NoStop}%
\bibitem [{\citenamefont {Chatterjee}\ \emph {et~al.}(2021)\citenamefont
  {Chatterjee}, \citenamefont {Wang}, \citenamefont {Berg},\ and\ \citenamefont
  {Zaletel}}]{CWBZ21}%
  \BibitemOpen
  \bibfield  {author} {\bibinfo {author} {\bibfnamefont {S.}~\bibnamefont
  {Chatterjee}}, \bibinfo {author} {\bibfnamefont {T.}~\bibnamefont {Wang}},
  \bibinfo {author} {\bibfnamefont {E.}~\bibnamefont {Berg}},\ and\ \bibinfo
  {author} {\bibfnamefont {M.~P.}\ \bibnamefont {Zaletel}},\ }\bibfield
  {title} {\bibinfo {title} {Inter-valley coherent order and isospin
  fluctuation mediated superconductivity in rhombohedral trilayer graphene},\
  }\href@noop {} {\  (\bibinfo {year} {2021})},\ \Eprint
  {https://arxiv.org/abs/arXiv:2109.00002} {arXiv:2109.00002} \BibitemShut
  {NoStop}%
\bibitem [{\citenamefont {McClure}(1969)}]{M69}%
  \BibitemOpen
  \bibfield  {author} {\bibinfo {author} {\bibfnamefont {J.~W.}\ \bibnamefont
  {McClure}},\ }\bibfield  {title} {\bibinfo {title} {Electron energy band
  structure and electronic properties of rhombohedral graphite},\ }\href@noop
  {} {\bibfield  {journal} {\bibinfo  {journal} {Carbon}\ }\textbf {\bibinfo
  {volume} {7}},\ \bibinfo {pages} {425} (\bibinfo {year} {1969})}\BibitemShut
  {NoStop}%
\bibitem [{\citenamefont {Dresselhaus}\ and\ \citenamefont
  {Dresselhaus}(2002)}]{DD02}%
  \BibitemOpen
  \bibfield  {author} {\bibinfo {author} {\bibfnamefont {M.~S.}\ \bibnamefont
  {Dresselhaus}}\ and\ \bibinfo {author} {\bibfnamefont {G.}~\bibnamefont
  {Dresselhaus}},\ }\bibfield  {title} {\bibinfo {title} {Intercalation
  compounds of graphite},\ }\href@noop {} {\bibfield  {journal} {\bibinfo
  {journal} {Adv. in Phys.}\ }\textbf {\bibinfo {volume} {51}},\ \bibinfo
  {pages} {1} (\bibinfo {year} {2002})}\BibitemShut {NoStop}%
\bibitem [{\citenamefont {Arovas}\ and\ \citenamefont {Guinea}(2008)}]{AG08}%
  \BibitemOpen
  \bibfield  {author} {\bibinfo {author} {\bibfnamefont {D.~P.}\ \bibnamefont
  {Arovas}}\ and\ \bibinfo {author} {\bibfnamefont {F.}~\bibnamefont
  {Guinea}},\ }\bibfield  {title} {\bibinfo {title} {Stacking faults, bound
  states, and quantum hall plateaus in crystalline graphite},\ }\href
  {https://doi.org/10.1103/PhysRevB.78.245416} {\bibfield  {journal} {\bibinfo
  {journal} {Phys. Rev. B}\ }\textbf {\bibinfo {volume} {78}},\ \bibinfo
  {pages} {245416} (\bibinfo {year} {2008})}\BibitemShut {NoStop}%
\bibitem [{\citenamefont {Zhang}\ \emph
  {et~al.}(2010{\natexlab{a}})\citenamefont {Zhang}, \citenamefont {Sahu},
  \citenamefont {Min},\ and\ \citenamefont {MacDonald}}]{ZSMM10}%
  \BibitemOpen
  \bibfield  {author} {\bibinfo {author} {\bibfnamefont {F.}~\bibnamefont
  {Zhang}}, \bibinfo {author} {\bibfnamefont {B.}~\bibnamefont {Sahu}},
  \bibinfo {author} {\bibfnamefont {H.}~\bibnamefont {Min}},\ and\ \bibinfo
  {author} {\bibfnamefont {A.~H.}\ \bibnamefont {MacDonald}},\ }\bibfield
  {title} {\bibinfo {title} {Band structure of $abc$-stacked graphene
  trilayers},\ }\href {https://doi.org/10.1103/PhysRevB.82.035409} {\bibfield
  {journal} {\bibinfo  {journal} {Phys. Rev. B}\ }\textbf {\bibinfo {volume}
  {82}},\ \bibinfo {pages} {035409} (\bibinfo {year}
  {2010}{\natexlab{a}})}\BibitemShut {NoStop}%
\bibitem [{\citenamefont {Koshino}(2010)}]{koshino_prb10}%
  \BibitemOpen
  \bibfield  {author} {\bibinfo {author} {\bibfnamefont {M.}~\bibnamefont
  {Koshino}},\ }\bibfield  {title} {\bibinfo {title} {Interlayer screening
  effect in graphene multilayers with $aba$ and $abc$ stacking},\ }\href
  {https://doi.org/10.1103/PhysRevB.81.125304} {\bibfield  {journal} {\bibinfo
  {journal} {Phys. Rev. B}\ }\textbf {\bibinfo {volume} {81}},\ \bibinfo
  {pages} {125304} (\bibinfo {year} {2010})}\BibitemShut {NoStop}%
\bibitem [{\citenamefont {Bao}\ \emph {et~al.}(2011)\citenamefont {Bao},
  \citenamefont {Jing}, \citenamefont {Velasco~Jr.}, \citenamefont {Lee},
  \citenamefont {Liu}, \citenamefont {Standley}, \citenamefont {Aykol},
  \citenamefont {Cronin}, \citenamefont {Smirnov}, \citenamefont {Koshino},
  \citenamefont {McCann}, \citenamefont {Bockrath},\ and\ \citenamefont
  {Lau}}]{Betal11}%
  \BibitemOpen
  \bibfield  {author} {\bibinfo {author} {\bibfnamefont {W.}~\bibnamefont
  {Bao}}, \bibinfo {author} {\bibfnamefont {L.}~\bibnamefont {Jing}}, \bibinfo
  {author} {\bibfnamefont {J.}~\bibnamefont {Velasco~Jr.}}, \bibinfo {author}
  {\bibfnamefont {Y.}~\bibnamefont {Lee}}, \bibinfo {author} {\bibfnamefont
  {D.}~\bibnamefont {Liu}, \bibfnamefont {G.and~Tran}}, \bibinfo {author}
  {\bibfnamefont {B.}~\bibnamefont {Standley}}, \bibinfo {author}
  {\bibfnamefont {M.}~\bibnamefont {Aykol}}, \bibinfo {author} {\bibfnamefont
  {S.~B.}\ \bibnamefont {Cronin}}, \bibinfo {author} {\bibfnamefont
  {D.}~\bibnamefont {Smirnov}}, \bibinfo {author} {\bibfnamefont
  {M.}~\bibnamefont {Koshino}}, \bibinfo {author} {\bibfnamefont
  {E.}~\bibnamefont {McCann}}, \bibinfo {author} {\bibfnamefont
  {M.}~\bibnamefont {Bockrath}},\ and\ \bibinfo {author} {\bibfnamefont
  {C.~N.}\ \bibnamefont {Lau}},\ }\bibfield  {title} {\bibinfo {title}
  {Stacking-dependent band gap and quantum transport in trilayer graphene},\
  }\href@noop {} {\bibfield  {journal} {\bibinfo  {journal} {Nature Phys.}\
  }\textbf {\bibinfo {volume} {7}},\ \bibinfo {pages} {948} (\bibinfo {year}
  {2011})}\BibitemShut {NoStop}%
\bibitem [{\citenamefont {Kopnin}\ \emph {et~al.}(2011)\citenamefont {Kopnin},
  \citenamefont {Heikkil\"a},\ and\ \citenamefont {Volovik}}]{KHV11}%
  \BibitemOpen
  \bibfield  {author} {\bibinfo {author} {\bibfnamefont {N.~B.}\ \bibnamefont
  {Kopnin}}, \bibinfo {author} {\bibfnamefont {T.~T.}\ \bibnamefont
  {Heikkil\"a}},\ and\ \bibinfo {author} {\bibfnamefont {G.~E.}\ \bibnamefont
  {Volovik}},\ }\bibfield  {title} {\bibinfo {title} {High-temperature surface
  superconductivity in topological flat-band systems},\ }\href
  {https://doi.org/10.1103/PhysRevB.83.220503} {\bibfield  {journal} {\bibinfo
  {journal} {Phys. Rev. B}\ }\textbf {\bibinfo {volume} {83}},\ \bibinfo
  {pages} {220503} (\bibinfo {year} {2011})}\BibitemShut {NoStop}%
\bibitem [{\citenamefont {Kopnin}\ \emph {et~al.}(2013)\citenamefont {Kopnin},
  \citenamefont {Ij\"as}, \citenamefont {Harju},\ and\ \citenamefont
  {Heikkil\"a}}]{KHH13}%
  \BibitemOpen
  \bibfield  {author} {\bibinfo {author} {\bibfnamefont {N.~B.}\ \bibnamefont
  {Kopnin}}, \bibinfo {author} {\bibfnamefont {M.}~\bibnamefont {Ij\"as}},
  \bibinfo {author} {\bibfnamefont {A.}~\bibnamefont {Harju}},\ and\ \bibinfo
  {author} {\bibfnamefont {T.~T.}\ \bibnamefont {Heikkil\"a}},\ }\bibfield
  {title} {\bibinfo {title} {High-temperature surface superconductivity in
  rhombohedral graphite},\ }\href {https://doi.org/10.1103/PhysRevB.87.140503}
  {\bibfield  {journal} {\bibinfo  {journal} {Phys. Rev. B}\ }\textbf {\bibinfo
  {volume} {87}},\ \bibinfo {pages} {140503} (\bibinfo {year}
  {2013})}\BibitemShut {NoStop}%
\bibitem [{\citenamefont {Lee}\ \emph {et~al.}(2014)\citenamefont {Lee},
  \citenamefont {Tran}, \citenamefont {Myhro}, \citenamefont {Velasco},
  \citenamefont {Gillgren}, \citenamefont {Lau}, \citenamefont {Barlas},
  \citenamefont {Poumirol}, \citenamefont {Smirnov},\ and\ \citenamefont
  {Guinea}}]{Letal14}%
  \BibitemOpen
  \bibfield  {author} {\bibinfo {author} {\bibfnamefont {Y.}~\bibnamefont
  {Lee}}, \bibinfo {author} {\bibfnamefont {D.}~\bibnamefont {Tran}}, \bibinfo
  {author} {\bibfnamefont {K.}~\bibnamefont {Myhro}}, \bibinfo {author}
  {\bibfnamefont {J.}~\bibnamefont {Velasco}}, \bibinfo {author} {\bibfnamefont
  {N.}~\bibnamefont {Gillgren}}, \bibinfo {author} {\bibfnamefont {C.~N.}\
  \bibnamefont {Lau}}, \bibinfo {author} {\bibfnamefont {Y.}~\bibnamefont
  {Barlas}}, \bibinfo {author} {\bibfnamefont {J.~M.}\ \bibnamefont
  {Poumirol}}, \bibinfo {author} {\bibfnamefont {D.}~\bibnamefont {Smirnov}},\
  and\ \bibinfo {author} {\bibfnamefont {F.}~\bibnamefont {Guinea}},\
  }\bibfield  {title} {\bibinfo {title} {Competition between spontaneous
  symmetry breaking and single-particle gaps in trilayer graphene},\
  }\href@noop {} {\bibfield  {journal} {\bibinfo  {journal} {Nature Comm.}\
  }\textbf {\bibinfo {volume} {5}},\ \bibinfo {pages} {5656} (\bibinfo {year}
  {2014})}\BibitemShut {NoStop}%
\bibitem [{\citenamefont {Pamuk}\ \emph {et~al.}(2017)\citenamefont {Pamuk},
  \citenamefont {Baima}, \citenamefont {Mauri},\ and\ \citenamefont
  {Calandra}}]{PBM17}%
  \BibitemOpen
  \bibfield  {author} {\bibinfo {author} {\bibfnamefont {B.}~\bibnamefont
  {Pamuk}}, \bibinfo {author} {\bibfnamefont {J.}~\bibnamefont {Baima}},
  \bibinfo {author} {\bibfnamefont {F.}~\bibnamefont {Mauri}},\ and\ \bibinfo
  {author} {\bibfnamefont {M.}~\bibnamefont {Calandra}},\ }\bibfield  {title}
  {\bibinfo {title} {Magnetic gap opening in rhombohedral-stacked multilayer
  graphene from first principles},\ }\href
  {https://doi.org/10.1103/PhysRevB.95.075422} {\bibfield  {journal} {\bibinfo
  {journal} {Phys. Rev. B}\ }\textbf {\bibinfo {volume} {95}},\ \bibinfo
  {pages} {075422} (\bibinfo {year} {2017})}\BibitemShut {NoStop}%
\bibitem [{\citenamefont {Chen}\ \emph
  {et~al.}(2019{\natexlab{a}})\citenamefont {Chen}, \citenamefont {Jiang},
  \citenamefont {Wu}, \citenamefont {Lyu}, \citenamefont {Li}, \citenamefont
  {Chittari}, \citenamefont {Watanabe}, \citenamefont {Taniguchi},
  \citenamefont {Shi}, \citenamefont {Jung}, \citenamefont {Zhang},\ and\
  \citenamefont {Wang}}]{Cetal19}%
  \BibitemOpen
  \bibfield  {author} {\bibinfo {author} {\bibfnamefont {G.}~\bibnamefont
  {Chen}}, \bibinfo {author} {\bibfnamefont {L.}~\bibnamefont {Jiang}},
  \bibinfo {author} {\bibfnamefont {S.}~\bibnamefont {Wu}}, \bibinfo {author}
  {\bibfnamefont {B.}~\bibnamefont {Lyu}}, \bibinfo {author} {\bibfnamefont
  {H.}~\bibnamefont {Li}}, \bibinfo {author} {\bibfnamefont {B.~L.}\
  \bibnamefont {Chittari}}, \bibinfo {author} {\bibfnamefont {K.}~\bibnamefont
  {Watanabe}}, \bibinfo {author} {\bibfnamefont {T.}~\bibnamefont {Taniguchi}},
  \bibinfo {author} {\bibfnamefont {Z.}~\bibnamefont {Shi}}, \bibinfo {author}
  {\bibfnamefont {J.}~\bibnamefont {Jung}}, \bibinfo {author} {\bibfnamefont
  {Y.}~\bibnamefont {Zhang}},\ and\ \bibinfo {author} {\bibfnamefont
  {F.}~\bibnamefont {Wang}},\ }\bibfield  {title} {\bibinfo {title} {Evidence
  of a gate-tunable mott insulator in a trilayer graphene moiré
  superlattice},\ }\href@noop {} {\bibfield  {journal} {\bibinfo  {journal}
  {Nature Phys.}\ }\textbf {\bibinfo {volume} {15}},\ \bibinfo {pages} {237}
  (\bibinfo {year} {2019}{\natexlab{a}})}\BibitemShut {NoStop}%
\bibitem [{\citenamefont {Lee}\ \emph {et~al.}(2019)\citenamefont {Lee},
  \citenamefont {Che}, \citenamefont {Velasco~Jr.}, \citenamefont {Tran},
  \citenamefont {Baima}, \citenamefont {Mauri}, \citenamefont {Calandra},
  \citenamefont {Bockrath},\ and\ \citenamefont {Lau}}]{Letal19}%
  \BibitemOpen
  \bibfield  {author} {\bibinfo {author} {\bibfnamefont {Y.}~\bibnamefont
  {Lee}}, \bibinfo {author} {\bibfnamefont {S.}~\bibnamefont {Che}}, \bibinfo
  {author} {\bibfnamefont {J.}~\bibnamefont {Velasco~Jr.}}, \bibinfo {author}
  {\bibfnamefont {D.}~\bibnamefont {Tran}}, \bibinfo {author} {\bibfnamefont
  {J.}~\bibnamefont {Baima}}, \bibinfo {author} {\bibfnamefont
  {F.}~\bibnamefont {Mauri}}, \bibinfo {author} {\bibfnamefont
  {M.}~\bibnamefont {Calandra}}, \bibinfo {author} {\bibfnamefont
  {M.}~\bibnamefont {Bockrath}},\ and\ \bibinfo {author} {\bibfnamefont
  {C.~N.}\ \bibnamefont {Lau}},\ }\bibfield  {title} {\bibinfo {title} {Gate
  tunable magnetism and giant magnetoresistance in abc-stacked few-layer
  graphene},\ }\href@noop {} {\  (\bibinfo {year} {2019})},\ \Eprint
  {https://arxiv.org/abs/1911.04450} {1911.04450} \BibitemShut {NoStop}%
\bibitem [{\citenamefont {Yin}\ \emph {et~al.}(2019)\citenamefont {Yin},
  \citenamefont {Shi}, \citenamefont {Li}, \citenamefont {Zhang}, \citenamefont
  {Guo},\ and\ \citenamefont {He}}]{yin_prl19}%
  \BibitemOpen
  \bibfield  {author} {\bibinfo {author} {\bibfnamefont {L.-J.}\ \bibnamefont
  {Yin}}, \bibinfo {author} {\bibfnamefont {L.-J.}\ \bibnamefont {Shi}},
  \bibinfo {author} {\bibfnamefont {S.-Y.}\ \bibnamefont {Li}}, \bibinfo
  {author} {\bibfnamefont {Y.}~\bibnamefont {Zhang}}, \bibinfo {author}
  {\bibfnamefont {Z.-H.}\ \bibnamefont {Guo}},\ and\ \bibinfo {author}
  {\bibfnamefont {L.}~\bibnamefont {He}},\ }\bibfield  {title} {\bibinfo
  {title} {High-magnetic-field tunneling spectra of $abc$-stacked trilayer
  graphene on graphite},\ }\href
  {https://doi.org/10.1103/PhysRevLett.122.146802} {\bibfield  {journal}
  {\bibinfo  {journal} {Phys. Rev. Lett.}\ }\textbf {\bibinfo {volume} {122}},\
  \bibinfo {pages} {146802} (\bibinfo {year} {2019})}\BibitemShut {NoStop}%
\bibitem [{\citenamefont {Chittari}\ \emph {et~al.}(2019)\citenamefont
  {Chittari}, \citenamefont {Chen}, \citenamefont {Zhang}, \citenamefont
  {Wang},\ and\ \citenamefont {Jung}}]{chittari_prl19}%
  \BibitemOpen
  \bibfield  {author} {\bibinfo {author} {\bibfnamefont {B.~L.}\ \bibnamefont
  {Chittari}}, \bibinfo {author} {\bibfnamefont {G.}~\bibnamefont {Chen}},
  \bibinfo {author} {\bibfnamefont {Y.}~\bibnamefont {Zhang}}, \bibinfo
  {author} {\bibfnamefont {F.}~\bibnamefont {Wang}},\ and\ \bibinfo {author}
  {\bibfnamefont {J.}~\bibnamefont {Jung}},\ }\bibfield  {title} {\bibinfo
  {title} {Gate-tunable topological flat bands in trilayer graphene
  boron-nitride moir\'e superlattices},\ }\href
  {https://doi.org/10.1103/PhysRevLett.122.016401} {\bibfield  {journal}
  {\bibinfo  {journal} {Phys. Rev. Lett.}\ }\textbf {\bibinfo {volume} {122}},\
  \bibinfo {pages} {016401} (\bibinfo {year} {2019})}\BibitemShut {NoStop}%
\bibitem [{\citenamefont {Chen}\ \emph
  {et~al.}(2019{\natexlab{b}})\citenamefont {Chen}, \citenamefont {Sharpe},
  \citenamefont {Gallagher}, \citenamefont {Rosen}, \citenamefont {Fox},
  \citenamefont {Jiang}, \citenamefont {Lyu}, \citenamefont {Li}, \citenamefont
  {Watanabe}, \citenamefont {Taniguchi}, \citenamefont {Jung}, \citenamefont
  {Shi}, \citenamefont {Goldhaber-Gordon}, \citenamefont {Zhang},\ and\
  \citenamefont {Wang}}]{Chen2019}%
  \BibitemOpen
  \bibfield  {author} {\bibinfo {author} {\bibfnamefont {G.}~\bibnamefont
  {Chen}}, \bibinfo {author} {\bibfnamefont {A.~L.}\ \bibnamefont {Sharpe}},
  \bibinfo {author} {\bibfnamefont {P.}~\bibnamefont {Gallagher}}, \bibinfo
  {author} {\bibfnamefont {I.~T.}\ \bibnamefont {Rosen}}, \bibinfo {author}
  {\bibfnamefont {E.~J.}\ \bibnamefont {Fox}}, \bibinfo {author} {\bibfnamefont
  {L.}~\bibnamefont {Jiang}}, \bibinfo {author} {\bibfnamefont
  {B.}~\bibnamefont {Lyu}}, \bibinfo {author} {\bibfnamefont {H.}~\bibnamefont
  {Li}}, \bibinfo {author} {\bibfnamefont {K.}~\bibnamefont {Watanabe}},
  \bibinfo {author} {\bibfnamefont {T.}~\bibnamefont {Taniguchi}}, \bibinfo
  {author} {\bibfnamefont {J.}~\bibnamefont {Jung}}, \bibinfo {author}
  {\bibfnamefont {Z.}~\bibnamefont {Shi}}, \bibinfo {author} {\bibfnamefont
  {D.}~\bibnamefont {Goldhaber-Gordon}}, \bibinfo {author} {\bibfnamefont
  {Y.}~\bibnamefont {Zhang}},\ and\ \bibinfo {author} {\bibfnamefont
  {F.}~\bibnamefont {Wang}},\ }\bibfield  {title} {\bibinfo {title}
  {{Signatures of tunable superconductivity in a trilayer graphene moir{\'{e}}
  superlattice}},\ }\href {https://doi.org/10.1038/s41586-019-1393-y}
  {\bibfield  {journal} {\bibinfo  {journal} {Nature}\ }\textbf {\bibinfo
  {volume} {572}},\ \bibinfo {pages} {215} (\bibinfo {year}
  {2019}{\natexlab{b}})},\ \Eprint {https://arxiv.org/abs/1901.04621}
  {arXiv:1901.04621} \BibitemShut {NoStop}%
\bibitem [{\citenamefont {Chen}\ \emph
  {et~al.}(2019{\natexlab{c}})\citenamefont {Chen}, \citenamefont {Sharpe},
  \citenamefont {Fox}, \citenamefont {Zhang}, \citenamefont {Wang},
  \citenamefont {Jiang}, \citenamefont {Lyu}, \citenamefont {Li}, \citenamefont
  {Watanabe}, \citenamefont {Taniguchi}, \citenamefont {Shi}, \citenamefont
  {Senthil}, \citenamefont {Goldhaber-Gordon}, \citenamefont {Zhang},\ and\
  \citenamefont {Wang}}]{Chen2020}%
  \BibitemOpen
  \bibfield  {author} {\bibinfo {author} {\bibfnamefont {G.}~\bibnamefont
  {Chen}}, \bibinfo {author} {\bibfnamefont {A.~L.}\ \bibnamefont {Sharpe}},
  \bibinfo {author} {\bibfnamefont {E.~J.}\ \bibnamefont {Fox}}, \bibinfo
  {author} {\bibfnamefont {Y.-H.~Y.}\ \bibnamefont {Zhang}}, \bibinfo {author}
  {\bibfnamefont {S.}~\bibnamefont {Wang}}, \bibinfo {author} {\bibfnamefont
  {L.}~\bibnamefont {Jiang}}, \bibinfo {author} {\bibfnamefont
  {B.}~\bibnamefont {Lyu}}, \bibinfo {author} {\bibfnamefont {H.}~\bibnamefont
  {Li}}, \bibinfo {author} {\bibfnamefont {K.}~\bibnamefont {Watanabe}},
  \bibinfo {author} {\bibfnamefont {T.}~\bibnamefont {Taniguchi}}, \bibinfo
  {author} {\bibfnamefont {Z.}~\bibnamefont {Shi}}, \bibinfo {author}
  {\bibfnamefont {T.}~\bibnamefont {Senthil}}, \bibinfo {author} {\bibfnamefont
  {D.}~\bibnamefont {Goldhaber-Gordon}}, \bibinfo {author} {\bibfnamefont
  {Y.-H.~Y.}\ \bibnamefont {Zhang}},\ and\ \bibinfo {author} {\bibfnamefont
  {F.}~\bibnamefont {Wang}},\ }\bibfield  {title} {\bibinfo {title} {{Tunable
  Correlated Chern Insulator and Ferromagnetism in Trilayer Graphene/Boron
  Nitride Moir\'e Superlattice}},\ }\href
  {https://doi.org/10.1038/s41586-020-2049-7} {\bibfield  {journal} {\bibinfo
  {journal} {Nature}\ }\textbf {\bibinfo {volume} {579}},\ \bibinfo {pages}
  {56} (\bibinfo {year} {2019}{\natexlab{c}})},\ \Eprint
  {https://arxiv.org/abs/1905.06535} {1905.06535} \BibitemShut {NoStop}%
\bibitem [{\citenamefont {Shi}\ \emph {et~al.}(2020{\natexlab{a}})\citenamefont
  {Shi}, \citenamefont {Xu}, \citenamefont {Yang}, \citenamefont {Slizovskiy},
  \citenamefont {Morozov}, \citenamefont {Son}, \citenamefont {Ozdemir},
  \citenamefont {Mullan}, \citenamefont {Barrier}, \citenamefont {Yin},
  \citenamefont {Berdyugin}, \citenamefont {Piot}, \citenamefont {Taniguchi},
  \citenamefont {Watanabe}, \citenamefont {Fal'ko}, \citenamefont {Novoselov},
  \citenamefont {Geim},\ and\ \citenamefont {Mishchenko}}]{Setal20}%
  \BibitemOpen
  \bibfield  {author} {\bibinfo {author} {\bibfnamefont {Y.}~\bibnamefont
  {Shi}}, \bibinfo {author} {\bibfnamefont {S.}~\bibnamefont {Xu}}, \bibinfo
  {author} {\bibfnamefont {Y.}~\bibnamefont {Yang}}, \bibinfo {author}
  {\bibfnamefont {S.}~\bibnamefont {Slizovskiy}}, \bibinfo {author}
  {\bibfnamefont {S.~V.}\ \bibnamefont {Morozov}}, \bibinfo {author}
  {\bibfnamefont {S.-K.}\ \bibnamefont {Son}}, \bibinfo {author} {\bibfnamefont
  {S.}~\bibnamefont {Ozdemir}}, \bibinfo {author} {\bibfnamefont
  {C.}~\bibnamefont {Mullan}}, \bibinfo {author} {\bibfnamefont
  {J.}~\bibnamefont {Barrier}}, \bibinfo {author} {\bibfnamefont
  {J.}~\bibnamefont {Yin}}, \bibinfo {author} {\bibfnamefont {A.~I.}\
  \bibnamefont {Berdyugin}}, \bibinfo {author} {\bibfnamefont {B.~A.}\
  \bibnamefont {Piot}}, \bibinfo {author} {\bibfnamefont {T.}~\bibnamefont
  {Taniguchi}}, \bibinfo {author} {\bibfnamefont {K.}~\bibnamefont {Watanabe}},
  \bibinfo {author} {\bibfnamefont {V.~I.}\ \bibnamefont {Fal'ko}}, \bibinfo
  {author} {\bibfnamefont {K.~S.}\ \bibnamefont {Novoselov}}, \bibinfo {author}
  {\bibfnamefont {A.~K.}\ \bibnamefont {Geim}},\ and\ \bibinfo {author}
  {\bibfnamefont {A.}~\bibnamefont {Mishchenko}},\ }\bibfield  {title}
  {\bibinfo {title} {Electronic phase separation in multilayer rhombohedral
  graphite},\ }\href@noop {} {\bibfield  {journal} {\bibinfo  {journal}
  {Nature}\ }\textbf {\bibinfo {volume} {584}},\ \bibinfo {pages} {210}
  (\bibinfo {year} {2020}{\natexlab{a}})}\BibitemShut {NoStop}%
\bibitem [{\citenamefont {{Zhou}}\ \emph
  {et~al.}(2021{\natexlab{b}})\citenamefont {{Zhou}}, \citenamefont {{Xie}},
  \citenamefont {{Ghazaryan}}, \citenamefont {{Holder}}, \citenamefont
  {{Ehrets}}, \citenamefont {{Spanton}}, \citenamefont {{Taniguchi}},
  \citenamefont {{Watanabe}}, \citenamefont {{Berg}}, \citenamefont
  {{Serbyn}},\ and\ \citenamefont {{Young}}}]{zhou_cm21_bis}%
  \BibitemOpen
  \bibfield  {author} {\bibinfo {author} {\bibfnamefont {H.}~\bibnamefont
  {{Zhou}}}, \bibinfo {author} {\bibfnamefont {T.}~\bibnamefont {{Xie}}},
  \bibinfo {author} {\bibfnamefont {A.}~\bibnamefont {{Ghazaryan}}}, \bibinfo
  {author} {\bibfnamefont {T.}~\bibnamefont {{Holder}}}, \bibinfo {author}
  {\bibfnamefont {J.~R.}\ \bibnamefont {{Ehrets}}}, \bibinfo {author}
  {\bibfnamefont {E.~M.}\ \bibnamefont {{Spanton}}}, \bibinfo {author}
  {\bibfnamefont {T.}~\bibnamefont {{Taniguchi}}}, \bibinfo {author}
  {\bibfnamefont {K.}~\bibnamefont {{Watanabe}}}, \bibinfo {author}
  {\bibfnamefont {E.}~\bibnamefont {{Berg}}}, \bibinfo {author} {\bibfnamefont
  {M.}~\bibnamefont {{Serbyn}}},\ and\ \bibinfo {author} {\bibfnamefont
  {A.~F.}\ \bibnamefont {{Young}}},\ }\bibfield  {title} {\bibinfo {title}
  {{Half and quarter metals in rhombohedral trilayer graphene}},\ }\href@noop
  {} {\bibfield  {journal} {\bibinfo  {journal} {arXiv e-prints}\ ,\ \bibinfo
  {eid} {arXiv:2104.00653}} (\bibinfo {year} {2021}{\natexlab{b}})},\ \Eprint
  {https://arxiv.org/abs/2104.00653} {arXiv:2104.00653 [cond-mat.mes-hall]}
  \BibitemShut {NoStop}%
\bibitem [{\citenamefont {Armitage}\ \emph {et~al.}(2018)\citenamefont
  {Armitage}, \citenamefont {Mele},\ and\ \citenamefont {Vishwanath}}]{AMV18}%
  \BibitemOpen
  \bibfield  {author} {\bibinfo {author} {\bibfnamefont {N.~P.}\ \bibnamefont
  {Armitage}}, \bibinfo {author} {\bibfnamefont {E.~J.}\ \bibnamefont {Mele}},\
  and\ \bibinfo {author} {\bibfnamefont {A.}~\bibnamefont {Vishwanath}},\
  }\bibfield  {title} {\bibinfo {title} {Weyl and dirac semimetals in
  three-dimensional solids},\ }\href
  {https://doi.org/10.1103/RevModPhys.90.015001} {\bibfield  {journal}
  {\bibinfo  {journal} {Rev. Mod. Phys.}\ }\textbf {\bibinfo {volume} {90}},\
  \bibinfo {pages} {015001} (\bibinfo {year} {2018})}\BibitemShut {NoStop}%
\bibitem [{\citenamefont {Cao}\ \emph {et~al.}(2018{\natexlab{a}})\citenamefont
  {Cao}, \citenamefont {Fatemi}, \citenamefont {Demir}, \citenamefont {Fang},
  \citenamefont {Tomarken}, \citenamefont {Luo}, \citenamefont
  {Sanchez-Yamagishi}, \citenamefont {Watanabe}, \citenamefont {Taniguchi},
  \citenamefont {Kaxiras}, \citenamefont {Ashoori},\ and\ \citenamefont
  {Jarillo-Herrero}}]{Cao2018}%
  \BibitemOpen
  \bibfield  {author} {\bibinfo {author} {\bibfnamefont {Y.}~\bibnamefont
  {Cao}}, \bibinfo {author} {\bibfnamefont {V.}~\bibnamefont {Fatemi}},
  \bibinfo {author} {\bibfnamefont {A.}~\bibnamefont {Demir}}, \bibinfo
  {author} {\bibfnamefont {S.}~\bibnamefont {Fang}}, \bibinfo {author}
  {\bibfnamefont {S.~L.}\ \bibnamefont {Tomarken}}, \bibinfo {author}
  {\bibfnamefont {J.~Y.}\ \bibnamefont {Luo}}, \bibinfo {author} {\bibfnamefont
  {J.~D.}\ \bibnamefont {Sanchez-Yamagishi}}, \bibinfo {author} {\bibfnamefont
  {K.}~\bibnamefont {Watanabe}}, \bibinfo {author} {\bibfnamefont
  {T.}~\bibnamefont {Taniguchi}}, \bibinfo {author} {\bibfnamefont
  {E.}~\bibnamefont {Kaxiras}}, \bibinfo {author} {\bibfnamefont {R.~C.}\
  \bibnamefont {Ashoori}},\ and\ \bibinfo {author} {\bibfnamefont
  {P.}~\bibnamefont {Jarillo-Herrero}},\ }\bibfield  {title} {\bibinfo {title}
  {{Correlated insulator behaviour at half-filling in magic-angle graphene
  superlattices}},\ }\href {https://doi.org/10.1038/nature26154} {\bibfield
  {journal} {\bibinfo  {journal} {Nature}\ }\textbf {\bibinfo {volume} {556}},\
  \bibinfo {pages} {80} (\bibinfo {year} {2018}{\natexlab{a}})},\ \Eprint
  {https://arxiv.org/abs/1802.00553} {arXiv:1802.00553} \BibitemShut {NoStop}%
\bibitem [{\citenamefont {Cao}\ \emph {et~al.}(2018{\natexlab{b}})\citenamefont
  {Cao}, \citenamefont {Fatemi}, \citenamefont {Fang}, \citenamefont
  {Watanabe}, \citenamefont {Taniguchi}, \citenamefont {Kaxiras},\ and\
  \citenamefont {Jarillo-Herrero}}]{Cao2018_bis}%
  \BibitemOpen
  \bibfield  {author} {\bibinfo {author} {\bibfnamefont {Y.}~\bibnamefont
  {Cao}}, \bibinfo {author} {\bibfnamefont {V.}~\bibnamefont {Fatemi}},
  \bibinfo {author} {\bibfnamefont {S.}~\bibnamefont {Fang}}, \bibinfo {author}
  {\bibfnamefont {K.}~\bibnamefont {Watanabe}}, \bibinfo {author}
  {\bibfnamefont {T.}~\bibnamefont {Taniguchi}}, \bibinfo {author}
  {\bibfnamefont {E.}~\bibnamefont {Kaxiras}},\ and\ \bibinfo {author}
  {\bibfnamefont {P.}~\bibnamefont {Jarillo-Herrero}},\ }\bibfield  {title}
  {\bibinfo {title} {{Unconventional superconductivity in magic-angle graphene
  superlattices}},\ }\href {https://doi.org/10.1038/nature26160} {\bibfield
  {journal} {\bibinfo  {journal} {Nature}\ }\textbf {\bibinfo {volume} {556}},\
  \bibinfo {pages} {43} (\bibinfo {year} {2018}{\natexlab{b}})}\BibitemShut
  {NoStop}%
\bibitem [{\citenamefont {Kohn}\ and\ \citenamefont
  {Luttinger}(1965)}]{KL_prl65}%
  \BibitemOpen
  \bibfield  {author} {\bibinfo {author} {\bibfnamefont {W.}~\bibnamefont
  {Kohn}}\ and\ \bibinfo {author} {\bibfnamefont {J.~M.}\ \bibnamefont
  {Luttinger}},\ }\bibfield  {title} {\bibinfo {title} {New mechanism for
  superconductivity},\ }\href {https://doi.org/10.1103/PhysRevLett.15.524}
  {\bibfield  {journal} {\bibinfo  {journal} {Phys. Rev. Lett.}\ }\textbf
  {\bibinfo {volume} {15}},\ \bibinfo {pages} {524} (\bibinfo {year}
  {1965})}\BibitemShut {NoStop}%
\bibitem [{\citenamefont {Cea}\ and\ \citenamefont {Guinea}(2021)}]{CG21}%
  \BibitemOpen
  \bibfield  {author} {\bibinfo {author} {\bibfnamefont {T.}~\bibnamefont
  {Cea}}\ and\ \bibinfo {author} {\bibfnamefont {F.}~\bibnamefont {Guinea}},\
  }\bibfield  {title} {\bibinfo {title} {Coulomb interaction, phonons, and
  superconductivity in twisted bilayer graphene},\ }\bibfield  {journal}
  {\bibinfo  {journal} {Proceedings of the National Academy of Sciences}\
  }\textbf {\bibinfo {volume} {118}},\ \href
  {https://doi.org/10.1073/pnas.2107874118} {10.1073/pnas.2107874118} (\bibinfo
  {year} {2021})\BibitemShut {NoStop}%
\bibitem [{\citenamefont {{Tien Phong}}\ \emph {et~al.}(2021)\citenamefont
  {{Tien Phong}}, \citenamefont {{Pantale{\'o}n}}, \citenamefont {{Cea}},\ and\
  \citenamefont {{Guinea}}}]{phong_cm21}%
  \BibitemOpen
  \bibfield  {author} {\bibinfo {author} {\bibfnamefont {V.}~\bibnamefont
  {{Tien Phong}}}, \bibinfo {author} {\bibfnamefont {P.~A.}\ \bibnamefont
  {{Pantale{\'o}n}}}, \bibinfo {author} {\bibfnamefont {T.}~\bibnamefont
  {{Cea}}},\ and\ \bibinfo {author} {\bibfnamefont {F.}~\bibnamefont
  {{Guinea}}},\ }\bibfield  {title} {\bibinfo {title} {{Band Structure and
  Superconductivity in Twisted Trilayer Graphene}},\ }\href@noop {} {\bibfield
  {journal} {\bibinfo  {journal} {arXiv e-prints}\ ,\ \bibinfo {eid}
  {arXiv:2106.15573}} (\bibinfo {year} {2021})},\ \Eprint
  {https://arxiv.org/abs/2106.15573} {arXiv:2106.15573 [cond-mat.mes-hall]}
  \BibitemShut {NoStop}%
\bibitem [{\citenamefont {Zhang}\ \emph
  {et~al.}(2010{\natexlab{b}})\citenamefont {Zhang}, \citenamefont {Sahu},
  \citenamefont {Min},\ and\ \citenamefont {MacDonald}}]{zhang_prb10}%
  \BibitemOpen
  \bibfield  {author} {\bibinfo {author} {\bibfnamefont {F.}~\bibnamefont
  {Zhang}}, \bibinfo {author} {\bibfnamefont {B.}~\bibnamefont {Sahu}},
  \bibinfo {author} {\bibfnamefont {H.}~\bibnamefont {Min}},\ and\ \bibinfo
  {author} {\bibfnamefont {A.~H.}\ \bibnamefont {MacDonald}},\ }\bibfield
  {title} {\bibinfo {title} {Band structure of $abc$-stacked graphene
  trilayers},\ }\href {https://doi.org/10.1103/PhysRevB.82.035409} {\bibfield
  {journal} {\bibinfo  {journal} {Phys. Rev. B}\ }\textbf {\bibinfo {volume}
  {82}},\ \bibinfo {pages} {035409} (\bibinfo {year}
  {2010}{\natexlab{b}})}\BibitemShut {NoStop}%
\bibitem [{\citenamefont {Zibrov}\ \emph {et~al.}(2018)\citenamefont {Zibrov},
  \citenamefont {Rao}, \citenamefont {Kometter}, \citenamefont {Spanton},
  \citenamefont {Li}, \citenamefont {Dean}, \citenamefont {Taniguchi},
  \citenamefont {Watanabe}, \citenamefont {Serbyn},\ and\ \citenamefont
  {Young}}]{zibrov_prl18}%
  \BibitemOpen
  \bibfield  {author} {\bibinfo {author} {\bibfnamefont {A.~A.}\ \bibnamefont
  {Zibrov}}, \bibinfo {author} {\bibfnamefont {P.}~\bibnamefont {Rao}},
  \bibinfo {author} {\bibfnamefont {C.}~\bibnamefont {Kometter}}, \bibinfo
  {author} {\bibfnamefont {E.~M.}\ \bibnamefont {Spanton}}, \bibinfo {author}
  {\bibfnamefont {J.~I.~A.}\ \bibnamefont {Li}}, \bibinfo {author}
  {\bibfnamefont {C.~R.}\ \bibnamefont {Dean}}, \bibinfo {author}
  {\bibfnamefont {T.}~\bibnamefont {Taniguchi}}, \bibinfo {author}
  {\bibfnamefont {K.}~\bibnamefont {Watanabe}}, \bibinfo {author}
  {\bibfnamefont {M.}~\bibnamefont {Serbyn}},\ and\ \bibinfo {author}
  {\bibfnamefont {A.~F.}\ \bibnamefont {Young}},\ }\bibfield  {title} {\bibinfo
  {title} {Emergent dirac gullies and gully-symmetry-breaking quantum hall
  states in $aba$ trilayer graphene},\ }\href
  {https://doi.org/10.1103/PhysRevLett.121.167601} {\bibfield  {journal}
  {\bibinfo  {journal} {Phys. Rev. Lett.}\ }\textbf {\bibinfo {volume} {121}},\
  \bibinfo {pages} {167601} (\bibinfo {year} {2018})}\BibitemShut {NoStop}%
\bibitem [{\citenamefont {Shi}\ \emph {et~al.}(2020{\natexlab{b}})\citenamefont
  {Shi}, \citenamefont {Xu}, \citenamefont {Yang}, \citenamefont {Slizovskiy},
  \citenamefont {Morozov}, \citenamefont {Son}, \citenamefont {Ozdemir},
  \citenamefont {Mullan}, \citenamefont {Barrier}, \citenamefont {Yin},
  \citenamefont {Berdyugin}, \citenamefont {Piot}, \citenamefont {Taniguchi},
  \citenamefont {Watanabe}, \citenamefont {Fal'ko}, \citenamefont {Novoselov},
  \citenamefont {Geim},\ and\ \citenamefont {Mishchenko}}]{shi_nat20}%
  \BibitemOpen
  \bibfield  {author} {\bibinfo {author} {\bibfnamefont {Y.}~\bibnamefont
  {Shi}}, \bibinfo {author} {\bibfnamefont {S.}~\bibnamefont {Xu}}, \bibinfo
  {author} {\bibfnamefont {Y.}~\bibnamefont {Yang}}, \bibinfo {author}
  {\bibfnamefont {S.}~\bibnamefont {Slizovskiy}}, \bibinfo {author}
  {\bibfnamefont {S.~V.}\ \bibnamefont {Morozov}}, \bibinfo {author}
  {\bibfnamefont {S.-K.}\ \bibnamefont {Son}}, \bibinfo {author} {\bibfnamefont
  {S.}~\bibnamefont {Ozdemir}}, \bibinfo {author} {\bibfnamefont
  {C.}~\bibnamefont {Mullan}}, \bibinfo {author} {\bibfnamefont
  {J.}~\bibnamefont {Barrier}}, \bibinfo {author} {\bibfnamefont
  {J.}~\bibnamefont {Yin}}, \bibinfo {author} {\bibfnamefont {A.~I.}\
  \bibnamefont {Berdyugin}}, \bibinfo {author} {\bibfnamefont {B.~A.}\
  \bibnamefont {Piot}}, \bibinfo {author} {\bibfnamefont {T.}~\bibnamefont
  {Taniguchi}}, \bibinfo {author} {\bibfnamefont {K.}~\bibnamefont {Watanabe}},
  \bibinfo {author} {\bibfnamefont {V.~I.}\ \bibnamefont {Fal'ko}}, \bibinfo
  {author} {\bibfnamefont {K.~S.}\ \bibnamefont {Novoselov}}, \bibinfo {author}
  {\bibfnamefont {A.~K.}\ \bibnamefont {Geim}},\ and\ \bibinfo {author}
  {\bibfnamefont {A.}~\bibnamefont {Mishchenko}},\ }\bibfield  {title}
  {\bibinfo {title} {Electronic phase separation in multilayer rhombohedral
  graphite},\ }\href {https://doi.org/10.1038/s41586-020-2568-2} {\bibfield
  {journal} {\bibinfo  {journal} {Nature}\ }\textbf {\bibinfo {volume} {584}},\
  \bibinfo {pages} {210} (\bibinfo {year} {2020}{\natexlab{b}})}\BibitemShut
  {NoStop}%
\bibitem [{\citenamefont {Wehling}\ \emph {et~al.}(2011)\citenamefont
  {Wehling}, \citenamefont {\ifmmode \mbox{\c{S}}\else \c{S}\fi{}a\ifmmode
  \mbox{\c{s}}\else \c{s}\fi{}\ifmmode \imath \else \i
  \fi{}o\ifmmode~\breve{g}\else \u{g}\fi{}lu}, \citenamefont {Friedrich},
  \citenamefont {Lichtenstein}, \citenamefont {Katsnelson},\ and\ \citenamefont
  {Bl\"ugel}}]{wehling_prl11}%
  \BibitemOpen
  \bibfield  {author} {\bibinfo {author} {\bibfnamefont {T.~O.}\ \bibnamefont
  {Wehling}}, \bibinfo {author} {\bibfnamefont {E.}~\bibnamefont {\ifmmode
  \mbox{\c{S}}\else \c{S}\fi{}a\ifmmode \mbox{\c{s}}\else \c{s}\fi{}\ifmmode
  \imath \else \i \fi{}o\ifmmode~\breve{g}\else \u{g}\fi{}lu}}, \bibinfo
  {author} {\bibfnamefont {C.}~\bibnamefont {Friedrich}}, \bibinfo {author}
  {\bibfnamefont {A.~I.}\ \bibnamefont {Lichtenstein}}, \bibinfo {author}
  {\bibfnamefont {M.~I.}\ \bibnamefont {Katsnelson}},\ and\ \bibinfo {author}
  {\bibfnamefont {S.}~\bibnamefont {Bl\"ugel}},\ }\bibfield  {title} {\bibinfo
  {title} {Strength of effective coulomb interactions in graphene and
  graphite},\ }\href {https://doi.org/10.1103/PhysRevLett.106.236805}
  {\bibfield  {journal} {\bibinfo  {journal} {Phys. Rev. Lett.}\ }\textbf
  {\bibinfo {volume} {106}},\ \bibinfo {pages} {236805} (\bibinfo {year}
  {2011})}\BibitemShut {NoStop}%
\bibitem [{hun()}]{hund}%
  \BibitemOpen
  \href@noop {} {}\bibinfo {note} {Note that we consider solely the Coulomb
  interaction associated to fluctuations of the total charge. The inclusion of
  a spin-dependent, inter-valley Hund coupling can allow for the existence of
  singlet solutions\cite{CWBZ21,GHMB21}.}\BibitemShut {Stop}%
\end{thebibliography}%

\clearpage
\onecolumngrid

\setcounter{equation}{0}
\setcounter{figure}{0}
\setcounter{table}{0}
\setcounter{page}{1}
\makeatletter
\renewcommand{\theequation}{S\arabic{equation}}
\renewcommand{\thefigure}{S\arabic{figure}}

\begin{center}
\Large {\textit{Supplementary Information for} \\ Superconductivity from Repulsive Interactions in Rhombohedral Trilayer Graphene: a Kohn-Luttinger-Like Mechanism}
\end{center}
\begin{center} 
Tommaso Cea, Pierre A. Pantale\'on, \foreignlanguage{vietnamese}{Võ Tiến Phong} and Francisco Guinea
\end{center}

In the main text, all numerical calculations are performed using a full six-band tight-binding model (TB). This captures accurately the evolution of the Fermi surface and associated van Hove singularities under the application of a perpendicular bias. However, this approach is numerically expensive because it requires  Bloch wavefunctions throughout the entire Brillouin zone. Knowing that the low-energy physics is dominated by only wavefunctions near the zone corners. It is often useful to project to just these regions in momentum space to obtain a continuum model. Here, we characterize the small differences between the TB model and its various continuum models. From the Hamiltonian in Eq.~(1) and following the procedure in Ref.~\cite{zhang_prb10}, the low-energy Hamiltonian in valley $\xi$ is given by  

\begin{equation}
H=H_{0}+H_{s}+H_{s'}+H_{t}+H_{\Delta}\label{eq: CModel Hamil}
\end{equation}
where 
\begin{align}
H_{0} & =\frac{v_{0}^{3}}{\gamma_{1}^{2}}\left(\begin{array}{cc}
0 & \left(\pi^{\ast}\right)^{3}\\
\pi^{3} & 0
\end{array}\right),\nonumber \\
H_{s} & =\left(\delta-\frac{2v_{0}v_{4}\boldsymbol{k}^{2}}{\gamma_{1}}\right)\sigma_{0},\nonumber \\
H_{s'} & =\Delta_{2}\left[1-3\left(\frac{v_{0}\boldsymbol{k}}{\gamma_{1}}\right)^{2}\right]\sigma_{0},\label{eq: CModel Terms}\\
H_{t} & =\left(\frac{\gamma_{2}}{2}-\frac{2v_{0}v_{3}\boldsymbol{k}^{2}}{\gamma_{1}}\right)\sigma_{x},\nonumber \\
H_{\Delta} & =\Delta_{1}\left[1-\left(\frac{v_{0}\boldsymbol{k}}{\gamma_{1}}\right)^{2}\right]\sigma_{z},\nonumber 
\end{align}
with $\pi= \xi k_{x}+i k_{y}$, $v_{i}=\frac{\sqrt{3}a\gamma_{i}}{2\hbar^{2}},$ $a=2.46$ $\textrm{Å},$ and $\boldsymbol k = \sqrt{k_x^2+k_y^2}$. The first term, $H_{0}$, is the simplest $ABC$ trilayer Hamiltonian with only nearest-neighbour interlayer hopping and dominates at larger values of momentum $\boldsymbol{k}$. The second $H_{s}$ term results from the weak coupling between the first and the third layer. The term $H_{s'}$ is non-zero if the potential of the middle layer deviates from the average potential in layers $1$ and $3$. The term $H_{t}$ is responsible for the trigonal warping in the band structure. The last term $H_{\Delta}$ takes into account the external electrostatic potential acting at the outermost layers. This term breaks inversion symmetry and opens a gap that is responsible of the flattening of the bands. 

\begin{figure}
\includegraphics[scale=0.25]{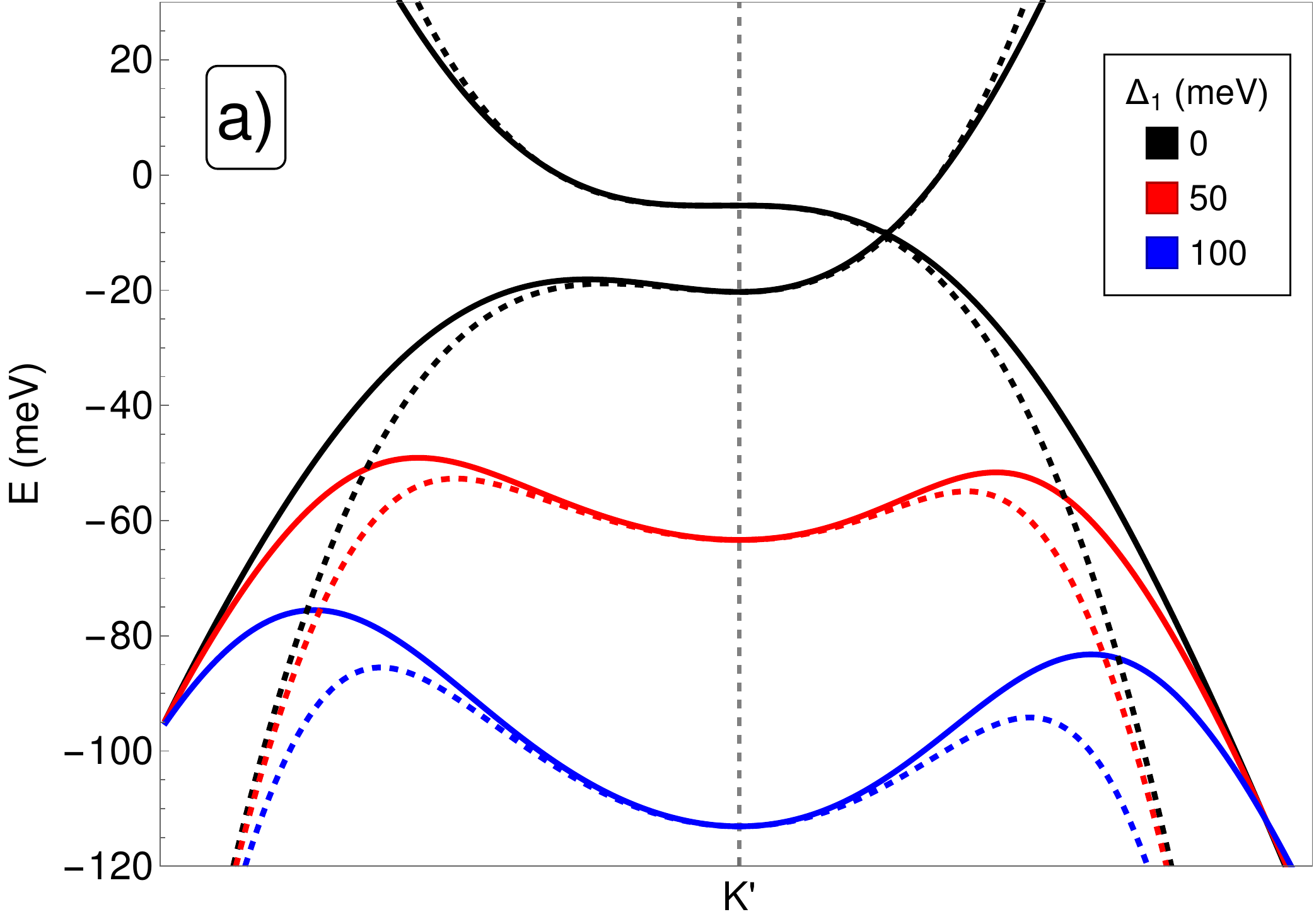} 
\includegraphics[scale=0.25]{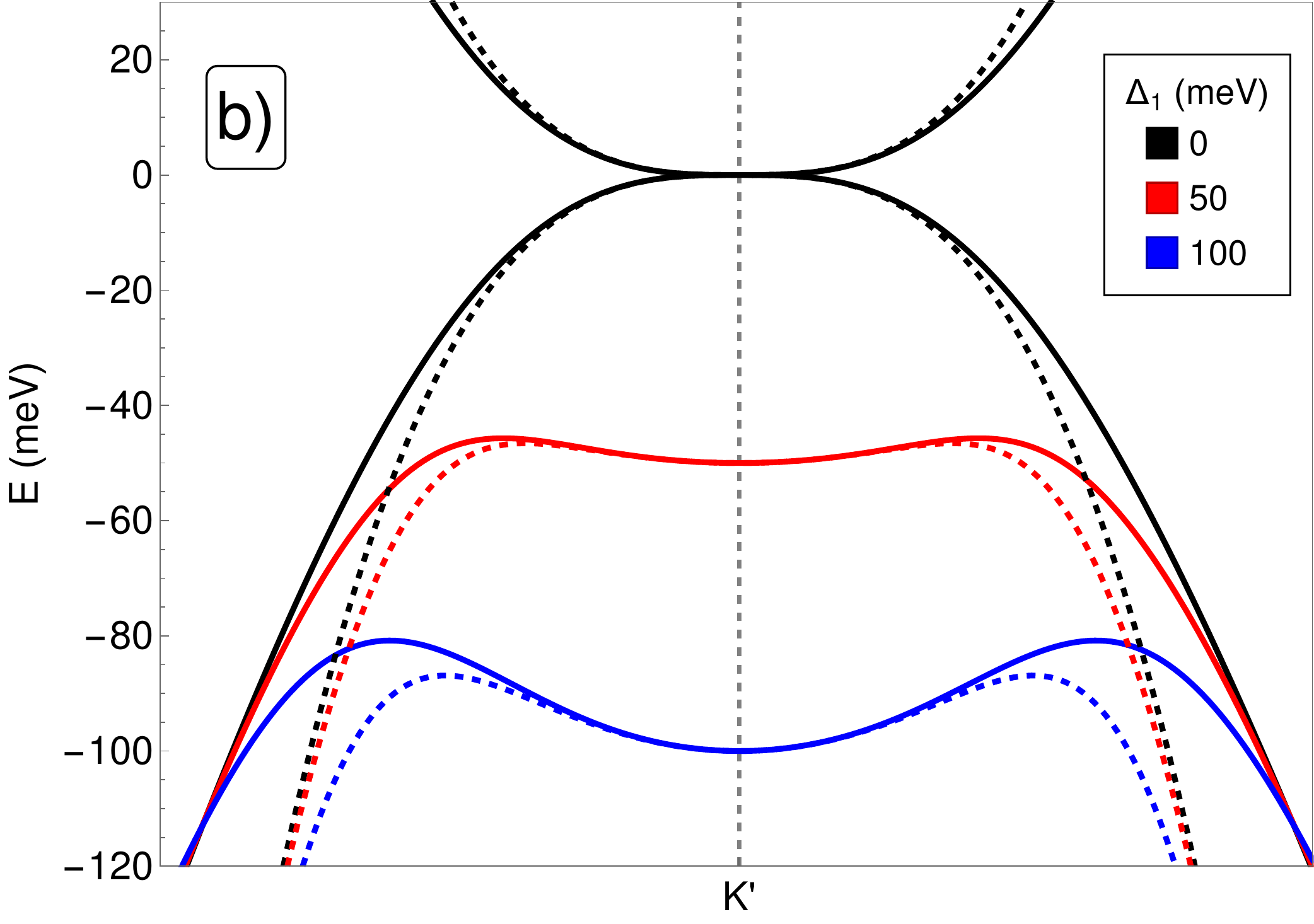} 
\includegraphics[scale=0.25]{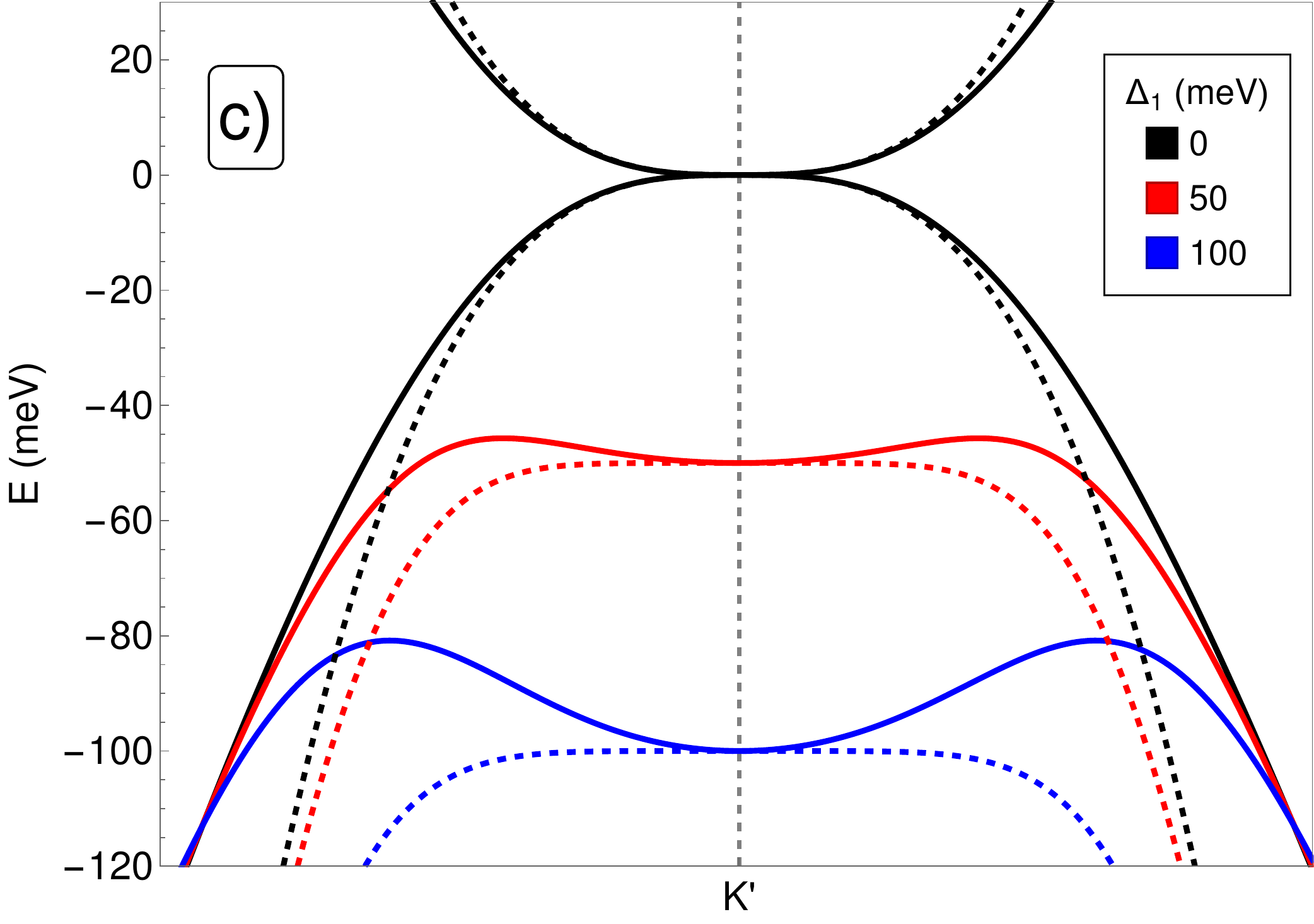} 
\caption{\textbf{Comparison of tight-binding and continuum models.} Band structure of ABC trilayer graphene for different values $\Delta_1$. Continuous lines are bands from the six-band full tight binding model of Eq.~(1), and dashed lines the continuum low-energy bands of Eq.~\eqref{eq: CModel Hamil} for different hopping values. In (a), all parameters are non-zero (as in Table 1 of the main text), in (b) $\gamma_0$, $\gamma_1$ and $\Delta_1$ are non-zero, and (c) is similar to (b) but without the quadratic term in $H_{\Delta}$. 
}
\label{fig: TBandCont}
\end{figure}

\begin{figure}
\includegraphics[scale=0.3]{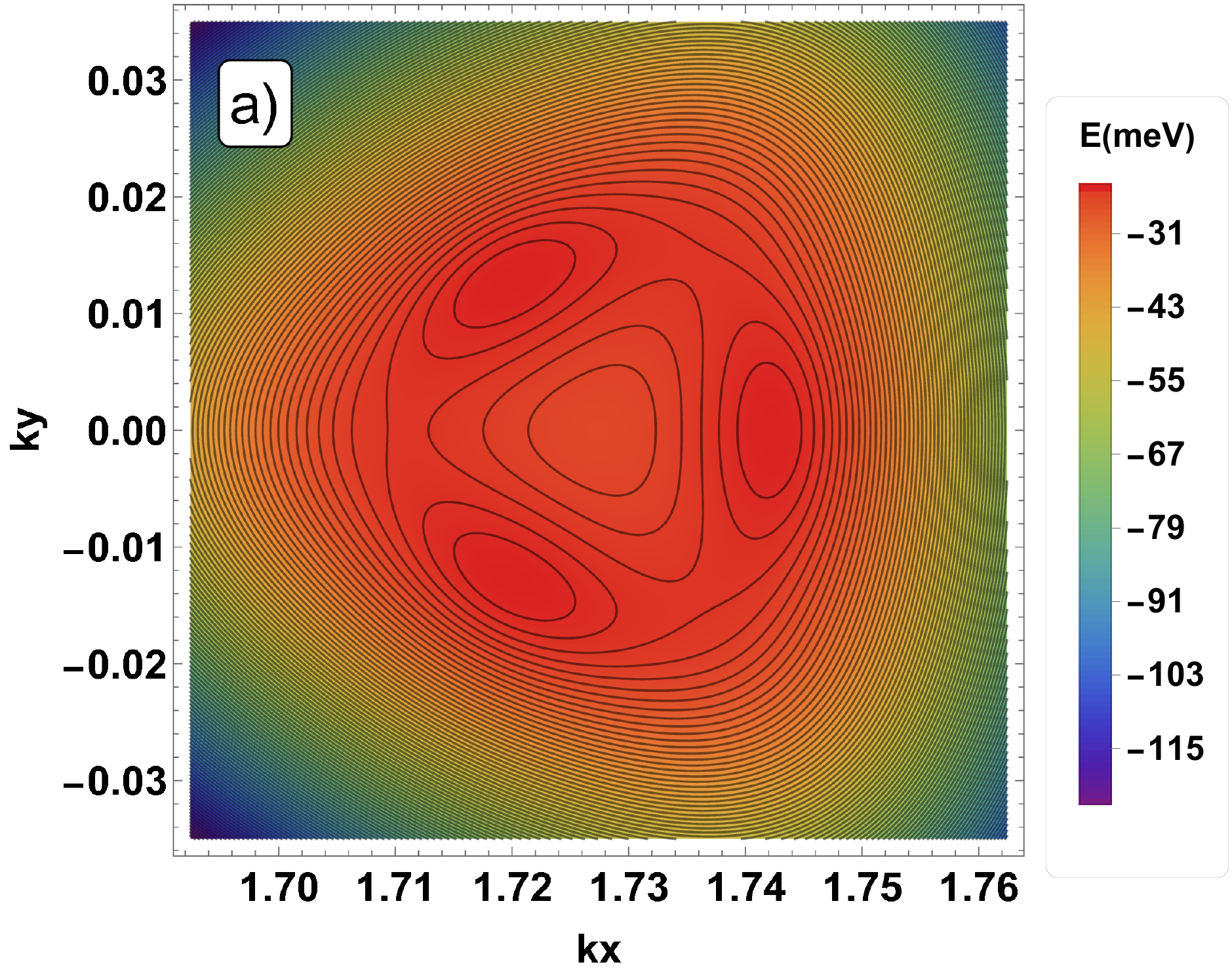} 
\includegraphics[scale=0.3]{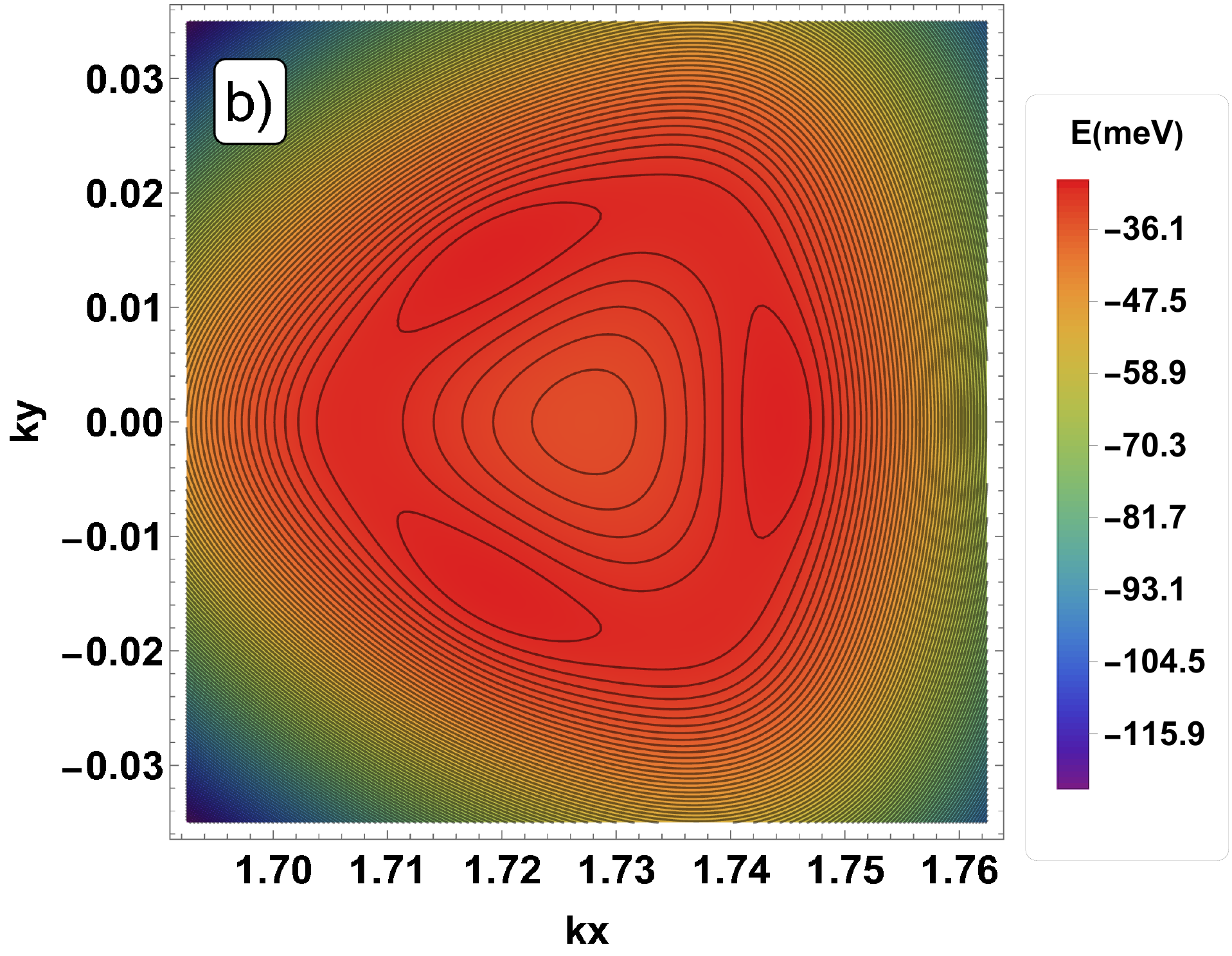} 
\includegraphics[scale=0.3]{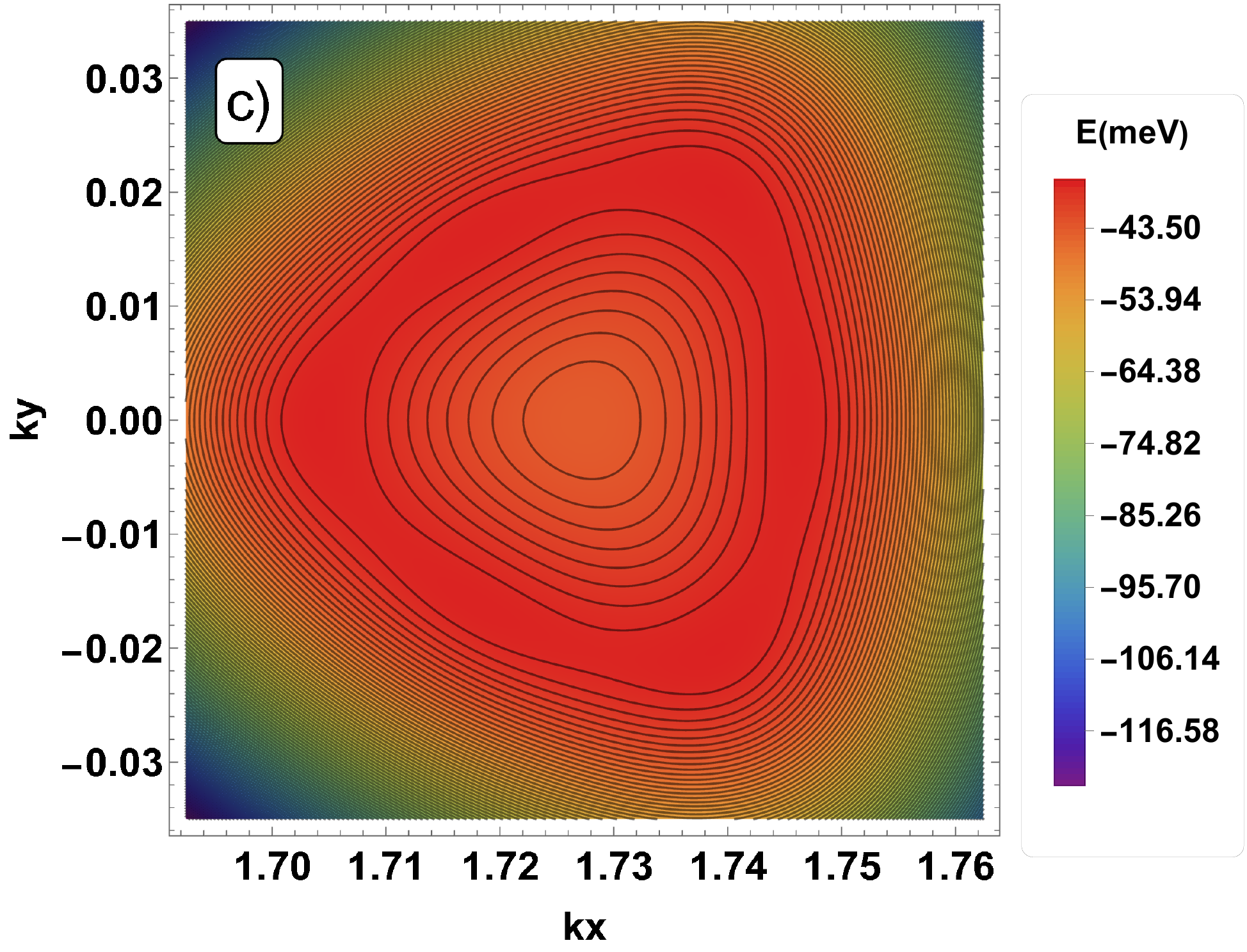} 
\includegraphics[scale=0.3]{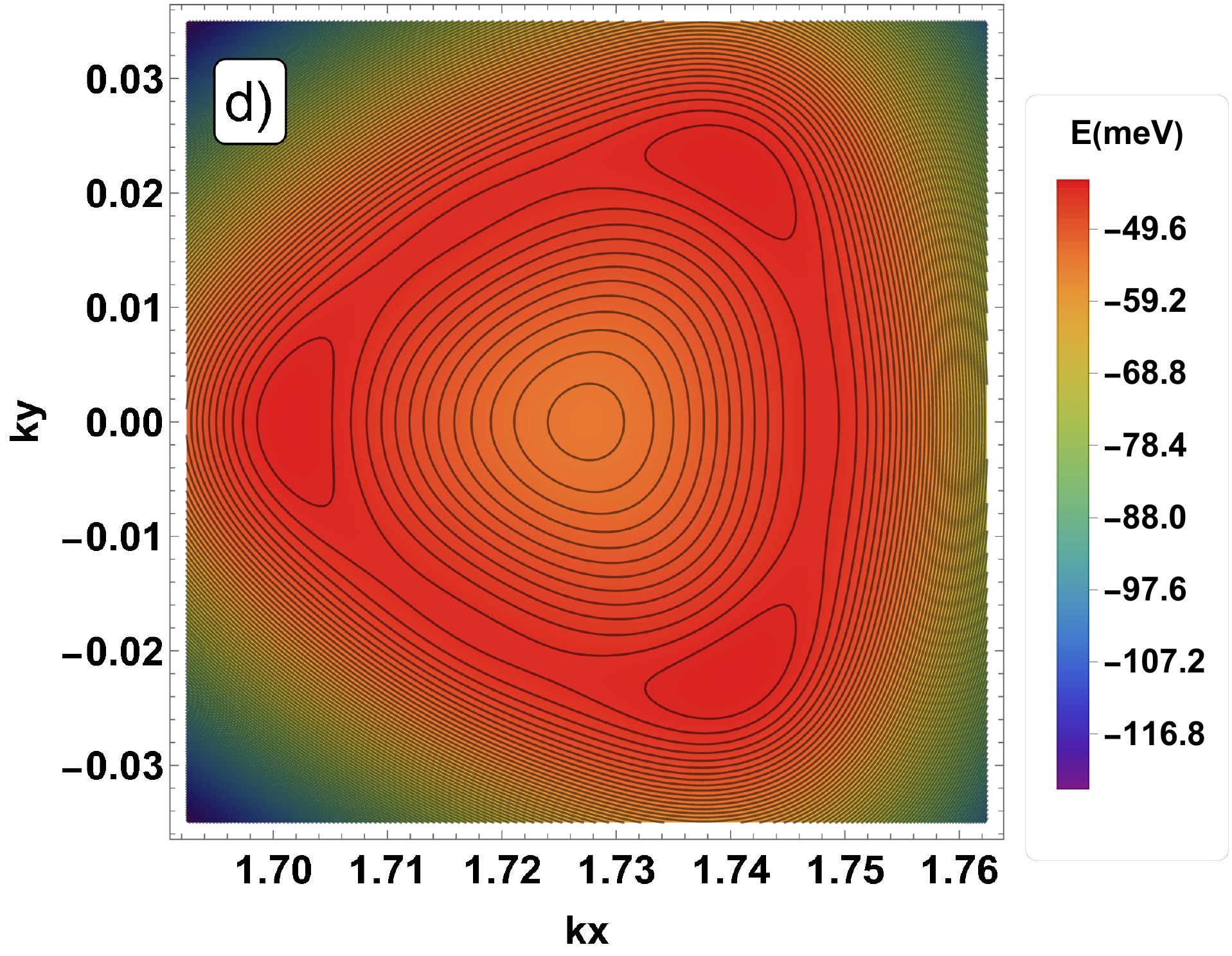} 
\includegraphics[scale=0.3]{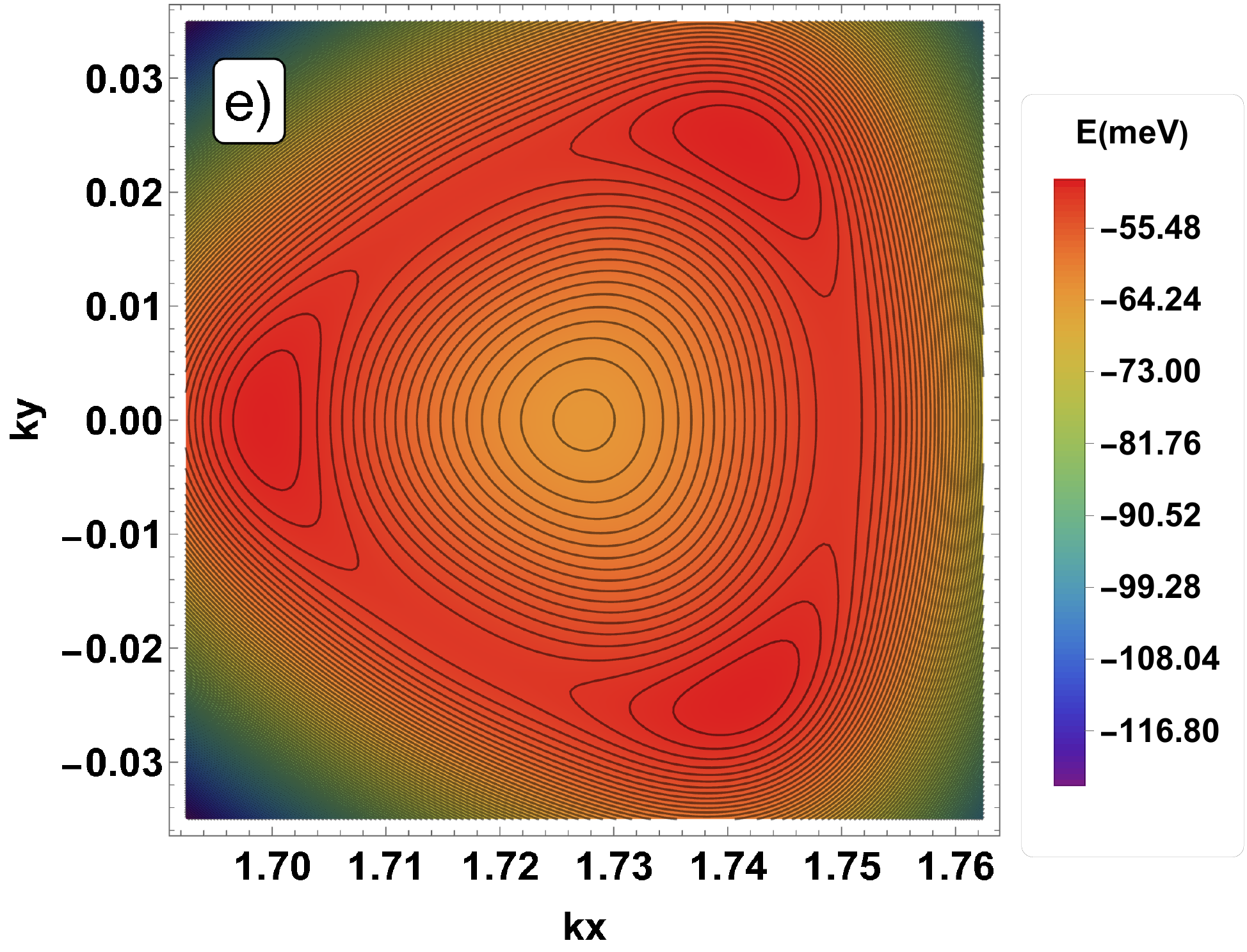} 
\includegraphics[scale=0.3]{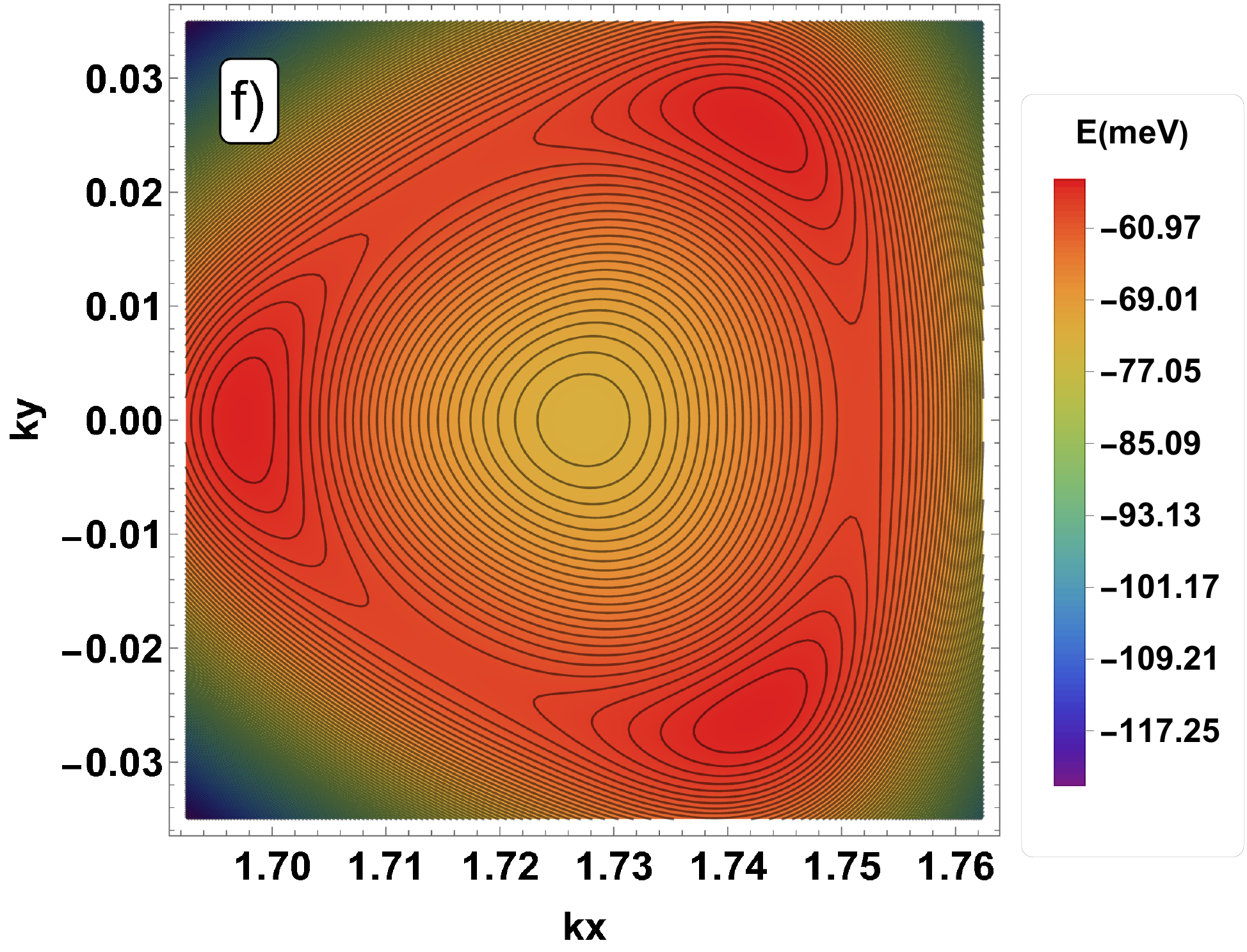} 
\caption{
\textbf{Evolution of the valence band structure near the $K'$ point.} Going from (a) to (f), the interlayer bias is increased as $\Delta_1={10,20,30,40,50}$ and $60$ meV respectively. Here, we use the full tight-binding model.}
\label{fig: TBDens}
\end{figure}

In Fig.~\ref{fig: TBandCont}, we compare the full TB model in Eq.~(1) with the continuum model in Eq.~\eqref{eq: CModel Hamil} for difference choices of parameters. In Fig~\ref{fig: TBandCont}(a), all parameters are non-zero, in this situation, as $\Delta_1$ increases, we distinguish in the energy spectra a parabolic-like dispersion almost centered at $K'$ point and two maxima. In the considered path, the maximum to the right corresponds to the Van Hove singularity in Fig.~5 of the main text. In Fig~\ref{fig: TBandCont}(b), we set to zero all the remote interlayer parameters except $\gamma_0$, $\gamma_1$ and $\Delta_1$. We find that the main features are preserved. However, in this case, we expect an increase in the DOS because both maxima are at the same energy. In Fig~\ref{fig: TBandCont}(c), we use the same conditions as in Fig~\ref{fig: TBandCont}(b), but we also remove the quadratic contribution in the last term in Eq.~\eqref{eq: CModel Hamil}, resulting in a term of the form $H_{\Delta}=\Delta_1 \sigma_{z}$. As a function of $\Delta_1,$ the bands near the Dirac point are always flat. If the filling is modified, there is no Lifshitz transition in the band structure. Figure~\ref{fig: TBDens} displays a density plot with the evolution of the valence bands for different values of the external displacement field $\Delta_1$ in the full tight-binding model. The gray lines in each figure are the the isoenergy contours.  

\end{document}